\documentclass[12pt]{article}
\usepackage{authblk}
\usepackage{geometry}
\usepackage{amsmath}
\usepackage{amssymb}
\usepackage{amsthm}
\usepackage{mathrsfs}
\usepackage{subcaption}
\usepackage{customcommands}
\usepackage{bm}
\usepackage{Nettastyle}
\usepackage{thmtools}
\usepackage{thm-restate}
\usepackage{physics}

\usepackage{tensor}
\usepackage{tcolorbox}

\usepackage[normalem]{ulem}

\newcommand{\CP}[1]{\ket{\sqrt{#1}}}
\newcommand{\CPS}[1]{\ket*{\sqrt{#1}}}

\preprint{MIT-CTP/5394}

\title{Canonical Purification of Evaporating Black Holes}

\author{Netta Engelhardt}
\author{and \AA{}smund Folkestad}
\emailAdd{engeln@mit.edu}
\emailAdd{afolkest@mit.edu}
\affiliation{Center for Theoretical Physics, Massachusetts Institute of Technology, \\Cambridge, MA 02139, USA}

\abstract{We show that the canonical purification of an evaporating black hole after the Page time consists of a short, connected, Lorentzian wormhole between two asymptotic boundaries, one of which is unitarily related to the radiation. This provides a quantitative and general realization of the predictions of ER=EPR in an evaporating black hole after the Page time; this further gives a standard AdS/CFT calculation of the entropy of the radiation (without modifications of the homology constraint). Before the Page time, the canonical purification consists of two disconnected, semiclassical black holes. From this, we construct two bipartite entangled holographic CFT states, with equal (and large) amount of entanglement, where  the semiclassical dual of one has a connected ERB and the other does not. From this example, we speculate that measures of multipartite entanglement may offer a more complete picture into the emergence of spacetime.
}

\begin{document}

\maketitle

\section{\label{sec:intro}Introduction}

The recent developments on the black hole information problem were catalyzed by a holographic calculation of the entropy of an AdS black hole evaporating into a reservoir~\cite{Pen19, AEMM}. Central to these results was the appearance of a novel quantum extremal surface (QES) $\chi$~\cite{EngWal14}, a stationary point of
\begin{equation}
    S_{\mathrm{gen}}[\chi]= \frac{\mathrm{Area}[\chi]}{4G_N} + S_{\mathrm{vN}}[\chi],
\end{equation}
where the second term above computes the von Neumann entropy on one side of the surface $\chi$. Contrary to expectations that QESs appear only in the vicinity of their classical counterparts (which compute von Neumann entropy in the classical setting~\cite{RyuTak06, HubRan07, FauLew13}), QESs crucially can appear even in spacetimes that have no classical extremal surfaces  -- such as black holes formed from collapse. See~\cite{AlmHar20} for a recent review. 

The von Neumann entropy is expected to play a central role in the emergence of spacetime (see work starting with \cite{Van09, Van10, CzeKar12, AlmDon14, DonHar16}). The closely related, but not identical, conjecture of ER=EPR proposes that (bipartite) entanglement between two parties can be geometrized by a spacetime connecting them \cite{MalSus13, Van13, VerVer13a}. (This was originally proposed as a way to make progress on the monogamy argument of the firewall problem~\cite{AMPS, AMPSS, Mat09} by encoding the black hole in the radiation itself, see~\cite{VerVer12, Sus13, Van13, PapRaj15} among others.)
This expectation is realized in many examples in AdS/CFT, most prominently in the case of the thermofield double on two
identical copies of a holographic CFT:
\begin{equation}\label{eq:TFD}
\begin{aligned}
    \ket{\text{TFD}} = \frac{ 1 }{ \sqrt{Z} }\sum_{n}e^{-\beta E_n/2}\ket{\tilde{n}} \ket{n}.
\end{aligned}
\end{equation}
For sufficiently high temperatures (and thus entanglement), this state is dual to a static black hole \cite{Mal01} with
an Einstein-Rosen bridge (ERB) connecting the asymptotic boundaries.
This is despite the fact that each individual term in the sum must be disconnected whenever a geometric dual exists. Since forming the sum creates both entanglement and
connection, it appears that entanglement is a crucial ingredient in connectedness.

There are several subtleties in the statements of ER=EPR and spacetime emergence from entanglement. When is the connection semiclassical? Without an independent
definition of a ``highly quantum'' wormhole, what is a sharp (falsifiable) formulation of ER=EPR?\footnote{Note that there is import in ER=EPR even in the absence of such a precise definition: if there is any sense in which the black hole interior is ``connected'' to the radiation, it is possible to act on the black hole interior by acting on the radiation. We thank J. Maldacena for discussion on this point.} In the specific context of AdS/CFT in the limit where the bulk is semiclassical, if a given bipartite CFT state (with a semiclassical dual) has a sufficiently large amount of entanglement , is it enough to
guarantee that a semiclassical spacetime that connects them is emergent?  Addressing these questions is a critical stepping stone towards understanding spacetime as an emergent phenomenon. 

In light of these unknowns and the novel developments starting with \cite{Pen19, AEMM}, let us ask the following: does the new QES after the Page time shed light on the
subtleties of ER=EPR and how or whether entanglement builds spacetime? 

A specific prediction of ER=EPR is that an old black hole should feature a semiclassical connection between the interior and the radiation. In the context of the recent developments on the holography of evaporating black holes, this question naively presents a puzzle: the island~\cite{AlmMah19} is not connected to the radiation via an ERB, suggesting a potential conflict with ER=EPR.\footnote{There is a sense in which the interface between the CFT and the bath provides a connection between the black hole and the radiation, but this is not a connection through a gravitational spacetime, and the connection can anyway be severed by turning on reflecting boundary conditions.} This is somewhat ameliorated by the doubly holographic model (see literature starting with~\cite{AlmMah19} in the context of the Page curve~\cite{Pag93a}), but the latter requires the bulk matter to be holographic with a classical bulk dual, which limits its range of applicability. Other connections include the Lewkowycz-Maldacena~\cite{LewMal13} style justifications for the novel QES via spacetime replica wormholes~\cite{PenShe19, AlmHar19}, where the latter may be a Euclidean avatar of ER=EPR; these non-factorizing geometries naturally present a new and exciting set of challenges~\cite{Col88, GidStr88, MalMao04}. Yet another possibility is turning on gravity in the reservoir -- without requiring double holography (see work starting with~\cite{GenKar20, AndPar21}) -- and subsequently testing whether an ERB forms dynamically. In~\cite{AndPar21}, it was shown that black holes evaporating into gravitating baths in certain toy models feature a Hawking-Page-like~\cite{HawPag83} transition after the Page time; of particular relevance is the work of~\cite{BalKar21, MiyUga21}, which started out with boundary conditions for two gravitating universes in JT gravity in the state~\eqref{eq:TFD} and showed that a Euclidean path integral preparation gave a dominant connected geometry for the computation of the von Neumann entropy. Yet another approach to connect the black hole to the radiation was made in \cite{BalKar21b}, relying crucially on an  end-of-the-world brane to cap off the black hole spacetime at $r > 0$. This allowed an additional spacetime representing a toy model of the radiation to be glued to the brane to build connection.

As advocated in~\cite{MarMax20b}, we would like to work fully in the Lorentzian bulk, in broad generality, and via operational definitions. Recall that the expectation expressed in~\cite{MalSus13} is not necessarily that a semiclassical connection exists in the original state, but that the application of a (high complexity) unitary to the radiation should result in a semiclassical wormhole connecting the old black hole to the radiation; that is, since the spacetime is not already semiclassically connected, a unitary on the radiation should render it so. We will therefore be concerned with finding such a unitary map, independently of whether the bath is gravitating or not. Our procedure will be entirely Lorentzian and fully generalizable to any number of dimensions, choice of UV completion (under assumption of the QES formula), and reasonable generic matter. 

We give a precise and general procedure that shows that there exists a unitary acting only on the radiation of an old black hole that converts the entanglement between an old black hole and its radiation into a semiclassical ERB. In the context of a two-sided black hole, a simple way to realize this expectation is by merging the radiation with the left black hole. However, prior to the discovery of the novel QES, this perspective was highly nontrivial in the context of a single-sided black hole, as there was no clear geometric division between the ``black hole'' system and the subset of the interior that could be modified by high complexity operations on the radiation. 

The novel QES elucidates this picture: after the Page time, there is a region behind the horizon that can be modified by acting exclusively on the microscopic state of the radiation $\rho_{\mathrm{RAD}}$. This is a key ingredient: the existence of a nontrivial QES after the Page time means that a Cauchy slice of the entanglement wedge of a single-sided black hole is extendible (even though the boundary Cauchy slice is inextendible). One may hope to apply an explicit map:
\begin{equation}
    \rho_{\mathrm{RAD}}\rightarrow U \rho_{\mathrm{RAD}}U^{\dagger},
\end{equation}
whose bulk dual introduces a boundary behind the QES, realizing precisely the post-Page time expectations that entanglement builds spacetime. Since the novel QES is a generic phenomenon for old black holes \cite{BouSha21} (possibly even away from AdS~\cite{CheGor20, HarJia20, Van20, BouSha21}), it may be possible to give an algorithmic prescription for the map $U$ that realizes ER=EPR in general old black holes. 

We give a prescription that shows that such a unitary exists using the so-called canonical purification\footnote{Our prescription is an indirect proof of existence: the canonical purification of the black hole maps unitarily to the radiation, but we do not know how to explicitly write down this unitary.}. A mixed state $\rho$ can be purified by doubling the Hilbert space on which it acts ${\cal H}\rightarrow {\cal H} \otimes {\cal H}$, and essentially entangling $\rho$ with its CPT conjugate. A familiar example of this procedure is the purification of the Gibbs ensemble via the thermofield double state. As is standard in the literature, we shall refer to the canonically purified state as $\ket{\sqrt{\rho}}$, and to the trace of $\ket{\sqrt{\rho}}$ over the original Hilbert space (on which $\rho$ acts) as $\widetilde{\rho}$. This construction will be reviewed at greater length in Sec.~\ref{sec:CanPur}.  

\begin{figure}[t]
     \centering
     \includegraphics[width=1\textwidth]{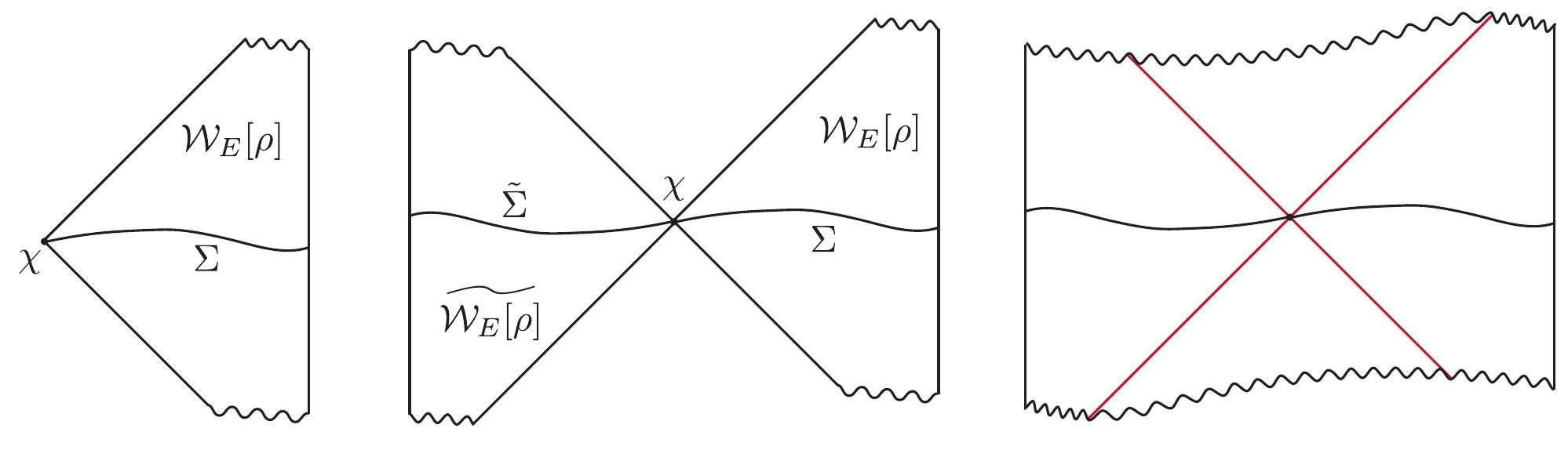}
     \caption{To the left we see the entanglement wedge of $\rho$, and in the center we see $\mathcal{W}_E[\rho]$ glued to its CPT conjugate across the QES/HRT surface $\chi$. To the right we display the final evolution of the data on $\Sigma \cup \tilde{\Sigma}$, giving the spacetime dual to $\CPS{\rho}$. Shockwaves (red) are present when including quantum corrections.}
     \label{fig:CP_sequence}
\end{figure} 

The canonical purification has a well-understood geometric dual.
Given a mixed CFT state $\rho$ with a classical bulk dual and entanglement wedge ${\cal W}_{E}[\rho]$, it was proposed
in~\cite{EngWal18} that $\widetilde{\rho}$ is dual to $\widetilde{{\cal W}_{E}[\rho]}$, the CPT-conjugate of ${\cal
W}_{E}[\rho]$, and the spacetime dual to the full canonically purified state $\CPS{\rho}$ is constructed by gluing
${\cal W}_{E}$ to $\widetilde{{\cal W}_{E}}$ along the HRT surface~\cite{HubRan07}. See Fig.~\ref{fig:CP_sequence}.~\cite{DutFau19} gave a path integral argument for this proposal (and further developed it in the context of the reflected entropy); the proposal was generalized to include bulk quantum corrections in~\cite{BouCha19}. When
quantum corrections are included, the bulk state is canonically purified as well, and the spacetime incurs a shock at
the gluing. The geometric dual to $\CPS{\rho}$ will also be reviewed in detail in Sec.~\ref{sec:CanPur}.

\begin{figure}[t]
     \centering
     \includegraphics[width=0.6\textwidth]{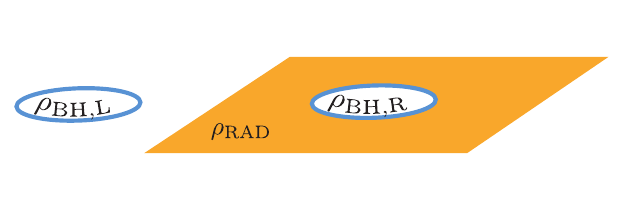}
     \caption{The three relevant non-gravitational systems in the microscopic picture of the evaporating two-sided black hole at a moment of time. The blue circles supports holographic CFTs, while the orange plane supports the reservoir system. }
     \label{fig:twoside_micro}
\end{figure}

\begin{figure}
     \centering
     \begin{subfigure}[b]{0.39\textwidth}
         \centering
         \includegraphics[width=0.9\textwidth]{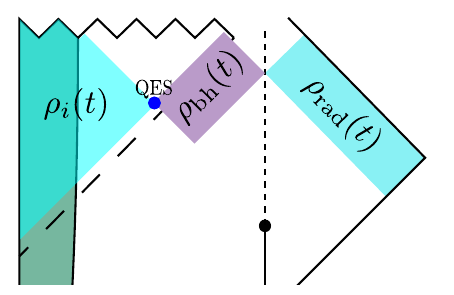}
         \caption{}
        
     \end{subfigure}
     \begin{subfigure}[b]{0.6\textwidth}
         \centering
         \includegraphics[width=0.7\textwidth]{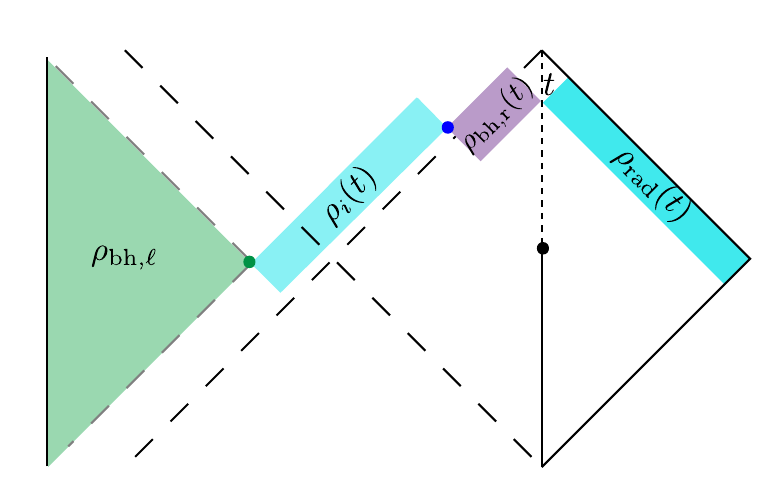}
         \caption{}
     \end{subfigure}
        \caption{The relevant subregions and states for evaporating one-sided (a) and two-sided (b) black holes in AdS.}
        \label{fig:bulkregions}
\end{figure}

We are now in position to explain our main construction. Consider an AdS black hole coupled to a reservoir, as described in~\cite{Pen19, AEMM}. 
There are two or three ``boundary'' subsystems of interest, depending on the whether the black hole is one- or two-sided~\footnote{We can similarly consider more boundaries; the story is qualitatively unchanged. Of course, if there are more than two CFT subsystems, there is an expectation that some multipartite entanglement will be important.}: the microscopic state of the radiation, with state $\rho_{\mathrm{RAD}}$, whose entropy follows the unitary Page curve~\cite{Pag93a}, and the state of the remaining black hole $\rho_{\mathrm{BH}}$; if there are two boundaries, we shall call the remaining
black hole states $\rho_{\mathrm{BH,L}}$ and $\rho_{\mathrm{BH,R}}$. For clarity we shall always take the right boundary to be coupled to the reservoir. This is illustrated in Fig.~\ref{fig:twoside_micro}. On the bulk side, we have three or four bulk regions: the non-gravitating, coarse-grained reservoir rad, whose state we shall denote $\rho_{\mathrm{rad}}$; the entanglement wedge of the old black hole, ${\cal W}_{E}[\rho_{\mathrm{BH}}]$, (or ${\cal W}_{E}[\rho_{\mathrm{BH,R}}]$ for two boundaries), whose state we shall denote $\rho_{\mathrm{bh}}$ (or $\rho_{\mathrm{bh,r}}$); the island~\cite{AlmMah19} ${i}$, whose state we shall denote as $\rho_{i}$; and if relevant, the left black hole ${\cal W}_{E}[\rho_{\mathrm{BH,L}}]$, whose state is denoted by $\rho_{\mathrm{bh, \ell}}$. These are illustrated in Fig.~\ref{fig:bulkregions}.

We build the canonical purification $\CPS{\rho_{\rm BH}(t)}$ of the black hole subsystem, working for now with a one-sided
black hole for pedagogical clarity. Before the Page time, at $t_1 < t_{\rm P}$, the QES is empty, and so gluing
$\mathcal{W}_E[\rho_{\rm BH}(t_1)]$ to its CPT conjugate across the QES amounts to simply introducing a second copy of the spacetime.\footnote{ $\widetilde{\mathcal{W}_E[\rho_{\rm BH}]}$ is introduced to replace the bath system as the purifier of $\rho_{\rm BH}$, so the bath system is removed rather than doubled.}

\begin{figure}
     \centering
     \includegraphics[width=0.8\textwidth]{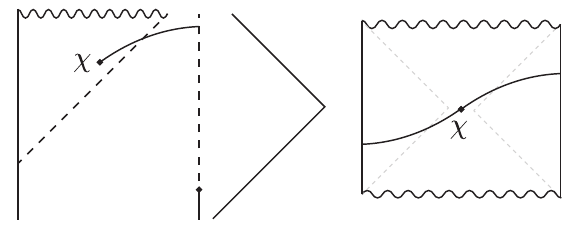}
     \caption{On the left we see the QES $\chi$ of an evaporating black hole for some time $t>t_P$. On the right, we see the canonical purification, dual to the state $\CPS{\rho_{BH}(t)}$.}
     \label{fig:CP_postpage}
\end{figure}

After the Page time, the dominant QES is nontrivial: CPT conjugation around the QES yields a \textit{connected} spacetime.~\footnote{This gluing procedure bears some resemblance to~\cite{PolStr94}, but the similarity is superficial.} Because the QES is close to the event horizon, this is a relatively short wormhole, reminiscent of (but not identical to) AdS-Schwarzschild. This is illustrated in Fig.~\ref{fig:CP_postpage}.  Since the QES for $\rho_{\mathrm{BH}}$ is identical to that of $\widetilde{\rho_{\mathrm{BH}}}$, the Page curve for the black hole immediately implies a Page curve for its canonical purification; since the von Neumann entropy is a unitary invariant (and purifications are unitarily-related), this also implies a Page curve for the radiation. Let us emphasize this point: by employing this unitary map, we circumvent any modification of the homology constraint in the QES formula: the standard QES formula applied to the canonical purification obeys the Page curve, which is a unitary invariant. The ERB that forms in the canonical purification after the Page time is a more conventional and more general manifestation of the connectivity that appears in the doubly holographic model. Because $\rho_{\rm RAD}$ and the canonical purification of $\rho_{\rm BH}$ are both purifications of $\rho_{\rm BH}$, they are related by a unitary. Thus this immediately shows that there exists a unitary acting exclusively on the post-Page radiation, which results in a connected spacetime.

We emphasize that while this story is consistent with earlier expectations that a black hole maximally mixed with its radiation should have a manifestation as an AdS-Schwarzschild-like geometry, this realization of that expectation is crucially reliant on the more recent understanding of the phenomenon of QESs far from classical extremal surfaces. 

It may be tempting at this point to conclude that the canonical purification is a good candidate for a precise realization of the `entanglement builds spacetime' paradigm: if the canonical purification of a density matrix with sufficient entanglement has a semiclassical bulk dual, then that bulk dual is connected. 

Here, however, we recall the case of the young single-sided black hole, whose canonical purification as described above is disconnected. Because the Page curve is non-monotonic, we may choose $t_{1}<t_P$ and $t_{2}>t_P$ such that $S_{\mathrm{vN}}[\rho_{\mathrm{BH}}(t_{1})]=S_{\mathrm{vN}}[\rho_{\mathrm{BH}}(t_{2})]$. However $\CPS{\rho_{\rm BH}(t_2)}$ gives a connected geometry, while $\CPS{\rho_{\rm BH}(t_1)}$ has a disconnected bulk dual \textit{even though they have the same von Neumann entropy}. This serves as a concrete example of a holographic semiclassical spacetime satisfying the standard QES prescription (with no modification of the homology constraint) in which the amount of fine-grained entropy cannot diagnose the emergence of spacetime connecting the bipartite subsystems. Let us  emphasize this point: while nonstandard generalizations of the QES formula that modify the homology constraint  -- i.e. the ``island'' formula -- may require a more liberal interpretation of spacetime emergence (e.g. in an additional dimension~\cite{AlmMah19} or via replica wormholes~\cite{PenShe19, AlmHar19}), here we have a conventional setup in which we may apply the standard QES formula (with the disconnected dominant topology in the replica trick), and we find that a large amount of bipartite entanglement cannot guarantee an emergent connected spacetime.

To summarize, our main technique relies on the following observation: that the canonical purification of an entanglement wedge with a nontrivial QES is connected, while the canonical purification of an entanglement wedge with an empty QES is disconnected. Our main purpose here consists of applications of this observation in a context where nontrivial QESs were previously unexpected (e.g. single-sided black holes formed from collapse). In those contexts, this property of the canonical purification allows us to (1) give an argument for the existence of the unitary $U_{\rm RAD}$ realizing ER=EPR, and (2) show that bipartite entanglement may fail to build spacetime even in bipartite states with semiclassical bulk duals. 

Furthermore, we show that obvious refinements to spacetime from bipartite entanglement conjecture, from reflected entropy to mutual information, complexity, or proximity to the maximally mixed state (as measured by $S_{\mathrm{thermal}}-S_{\mathrm{vN}}$) cannot be solely responsible for spacetime emergence. This requires a refinement of the general expectation (e.g.~\cite{JenSon14}) that a sufficient von Neumann entropy of bipartite holographic states necessarily builds spacetime between the subsystems. 

Under the assumption of entanglement wedge nesting and reconstruction, our results are robust against any unitary acting on the radiation: amount of entanglement is not a sufficient criterion for a semiclassical connection between the black hole and the (unitarily modified) radiation. It is simple to see why:  any Cauchy slice of the entanglement wedge of the young black hole is inextendible. Without modifying $\rho_{\mathrm{BH}}(t_{1})$ in some way, it is impossible to create a spacetime connection between it and any other spacetime.\footnote{Unless one adds additional spacetime dimensions, as in~\cite{AlmMah19}, or the semiclassical picture for $\rho_{\mathrm{BH}}$ is invalid.} 

This has somewhat different implications for ER=EPR and the specific question of spacetime emergence. There is at least one interpretation of ER=EPR that maintains its applicability to the black hole information problem and is consistent with our results: that a factorized unitary $U_{\rm RAD}\otimes U_{\rm BH}$ applied to the young black hole and its radiation $\ket{\Psi(t<t_{P})}$ can yield a connected geometry.\footnote{It is of course trivial that a nonfactorizing unitary can accomplish this.} This requires $U_{\rm BH}$ to change the topology of ${\cal W}_{E}[\rho_{\mathrm{BH}}]$, but there is no obvious obstacle to that. However, this interpretation poses a puzzle for spacetime emergence independent of the information problem considerations of ER=EPR. To understand this latter phenomenon, we would like to identify the property of a given quantum state (rather than some unitary equivalent) that generates spacetime. In the case of a young black hole, we have a bipartite state $\CPS{\rho(t<t_{P})}$ whose marginals have conventional semiclassical bulk duals with ${\cal O}(G_{N}^{-1})$ bipartite entanglement, and yet that entanglement fails to build spacetime between them. In the case of the canonically purified old black hole $\CPS{\rho(t>t_{P})}$, the same amount of bipartite entanglement successfully builds a connecting spacetime. It is immaterial that a factorized unitary could possibly be applied to make $\CPS{\rho(t<t_{P})}$ connected: our task, to understand what gives rise to an emergent spacetime in a given state, remains unsolved. In fact, this example makes it clear that whatever quantity is responsible, it cannot be a unitary invariant. This, along with other considerations that we will discuss in Sec.~\ref{sec:prePage}, is suggestive that \textit{multipartite} entanglement may play a crucial role in distinguishing between even \textit{bipartite} states with connected and disconnected semiclassical duals. We conclude with a discussion in Sec.~\ref{sec:discussion} of the shortfalls of various natural refinements and remaining possibilities, including the possible role of reconstruction within a code subspace of typical microstates.

\section{Holographic Canonical Purification}\label{sec:CanPur}
Let us begin with a brief review of the canonical purification of some mixed state $\rho$, for convenience chosen to be in a diagonal basis: \begin{equation}
    \rho = \sum\limits_{i} p_{i} |\rho_{i}\rangle \langle \rho_{i}|.
\end{equation}
We may obtain a pure state by doubling the Hilbert space and essentially duplicating the conjugated state by sending bras to kets:
\begin{equation}
    \CP{\rho} = \sum\limits_{i} \sqrt{p_{i} }\, |\rho_{i}\rangle |\tilde{\rho_{i}}\rangle,
\end{equation}
where $\ket{\tilde{\rho}_i}$  is the CPT conjugate of $\ket{\rho_i}$.
We will refer to the trace of $\CPS{\rho}$ over the original system as $\widetilde{\rho}$ for clarity.

A very familiar example of this procedure is the map from the Gibbs ensemble to the thermofield double
state:
\begin{equation}
    \rho_{\beta}=\frac{1}{Z}\sum e^{-\beta E_{n}}|n\rangle \langle n| \ \rightarrow \ket{\mathrm{TFD}}\equiv
    \CP{\rho_{\beta}} = \frac{1}{\sqrt{Z}}\sum e^{-\beta E_{n}/2}|n \rangle |\tilde{n} \rangle.
\end{equation}
The canonical purification may be thought of as the generalization of this procedure to more general, non-thermal states. The mathematically rigorous procedure, including the proof of existence and uniqueness, is simply an implementation of the GNS construction~\cite{GelNeu43, Seg47}.

In the classical bulk regime, the holographic dual of this purification was proposed~\cite{EngWal18} to be the CPT-conjugation of ${\cal W}_{E}[\rho]$ across the HRT surface. To be precise, we take any Cauchy slice of ${\cal W}_{E}[\rho]$ with its gravitational initial data $(\Sigma, h_{ab}, K_{ab})$ as well as initial data for any matter fields. In an abuse of notation, we will collectively refer to this entire set of initial data of ${\cal W}_{E}[\rho]$ as $\Sigma$. We then CPT-conjugate $\Sigma$ across the HRT surface, generating a maximally extended Cauchy slice for a complete, possibly multi-boundary spacetime.~\footnote{The spacetime will be multi-boundary whenever the HRT surface in question is nontrivial for a complete connected boundary. The procedure however can also be implemented for subregions.} This can be thought of as a gluing of $\Sigma$ to its CPT conjugate $\widetilde{\Sigma}$ across the HRT surface $\chi$. The initial data is then evolved via the standard Cauchy problem to generate the rest of the spacetime. This is illustrated in Fig.~\ref{fig:CP_sequence}.

The construction is straightforward largely due to the fact that the HRT surface by definition has vanishing mean
curvature $K^{a}=0$, which simplifies the junction conditions~\cite{BarIsr91, EngWal18} of $\Sigma$ with $\widetilde{\Sigma}$. Under the
inclusion of quantum corrections, the classical extremality condition $K^{a}=0$ is replaced by quantum
extremality. The  quantum corrected generalization of the holographic dual to the canonical purification naturally
requires CPT conjugation around the QES (and canonical purification of the bulk state)~\cite{BouCha19}. A QES, however, does not
generically satisfy $K^{a}=0$. Consequently, gluing $\Sigma$ to its CPT conjugate incurs a shock that corrects for the
mismatch between $K^{a}[\chi]$ on approach from $\Sigma$ and from $\widetilde{\Sigma}$. The precise form of the shock,
derived in~\cite{BouCha19}, is:\footnote{We must multiply by a factor of two, since we are CPT conjugating about a QES rather
than a quantum minimar, leading the two shocks of \cite{BouCha19} to coincide. }
\begin{equation}
    T_{kk}= \frac{ 1}{ \pi }\delta(\lambda) \frac{ \delta S_{\rm out}[V(\lambda)] }{ \delta V(\lambda)},
\end{equation}
where $k^a$ is a null vector orthogonal to $\chi$, $S_{\rm out}$ the von Neumann entropy of the bulk state outside of the QES, $\lambda$ the affine parameter of
$k^a$, and $V(\lambda)$ a function determining the location of a slice of the null congruence generated by $k^a$ -- see \cite{BouCha19} for details.
The protocol, including the expected location of the shocks, is illustrated in Fig.~\ref{fig:CP_sequence}.

Note that by definition, the canonical purification leaves the original state $\rho$ unaltered. Likewise, its bulk dual makes no modification to ${\cal W}_{E}[\rho]$. Thus any canonical purification of the boundary dual to an evaporating black hole will leave the black hole entanglement wedge unchanged.

\section{ER=EPR from the New QES}\label{sec:postPage}

\begin{figure}
     \centering
     \includegraphics[width=1\textwidth]{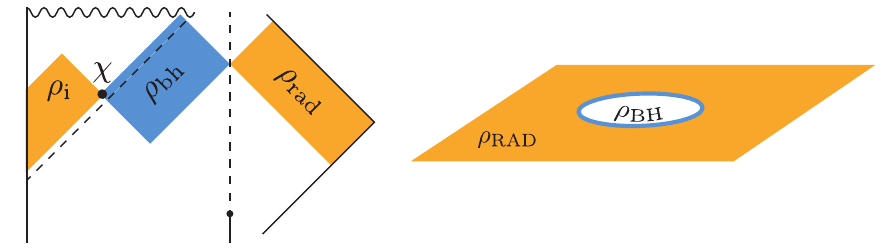}
     \caption{On the left we see the bulk picture of an AdS black hole evaporating into a reservoir, with $\chi$ the QES for some time $t>t_P$. On the right, we see the microscopic picture at the fixed time $t$. }
     \label{fig:regions}
\end{figure} 

As a central point of ER=EPR, \cite{MalSus13} conjectured that the Hawking radiation of an old black hole connects  to the black hole interior through a wormhole. While the wormhole may not be geometric, the proposal was that it can be made semiclassical by acting on the Hawking radiation with a unitary -- e.g. collapsing it into a second black hole~\cite{Van13}.

As explained in the introduction, we take
an asymptotically (say, one-sided) AdS black hole, which we shall initially take to be evaporating into a reservoir. Let $\ket{\Psi(t)}$ be the full state of the system on $\mathcal{H}_{\rm BH}\otimes \mathcal{H}_{\rm RAD}$ at boundary time $t>t_{P}$ and denote the reduced states as $\rho_{\rm BH}(t)$ and $\rho_{\rm RAD}(t)$.  See Fig.~\ref{fig:regions}.

The canonical purification $\CPS{\rho_{\rm BH}(t)} \in \mathcal{H}_{\rm BH}\otimes
\mathcal{H}_{\rm BH}$ after the
Page time is a new two-sided connected spacetime, which is CPT symmetric about $\chi(t)$; the bulk quantum state $\ket{\psi(t)}$ is the canonical purification of $\rho_{\rm bh}(t)$, the bulk quantum state in $\mathcal{W}_E[\rho_{\rm BH}(t)]$.  How would we diagnose connectivity in this case if we do not have access to the global geometry of spatial slices? 

Nonlinear quantities such as the mutual information have historically been proposed as appropriate diagnostic tools for
spacetime connectivity~\cite{Van09}. For instance, in a purely classical bulk, the mutual information of two subregions is
$\mathcal{O}(1/G_{N})$ only if their entanglement wedge is connected.  These ideas immediately break down once bulk
entanglement can make appreciable contributions to $S_{\mathrm{gen}}[\chi]$, as we will see in later sections.

It is fairly simple to give a specialized sufficient condition for connectivity of a class of low-complexity states: simply couple the two sides together via some Gao-Jafferis-Wall-style protocol~\cite{GaoJaf16} (in JT gravity  as in~\cite{MalSta17}).  If the QES is approximately on the horizon,  by introducing this coupling, we can make the wormhole traversable; this effect is easily detected by the near-boundary limit of bulk correlators~\cite{MalSta17}. In higher dimensions, under the assumption that a similar protocol as in~\cite{GaoJaf16} can be made to work in general, the same diagnostic may be used as a sufficient condition.  For the states that we consider in this paper this is actually enough to unambiguously diagnose connectivity.  If we execute this protocol for our post-Page time canonical purification, we will indeed recover connectivity for this spacetime.

As an aside, we may execute this protocol whenever the spacetime has no Python's lunch~\cite{BroGha19}: if the outermost QES is in fact the minimal QES. In that case, the intuition would be that the outermost (and in this case, minimal) QES lives on the bifurcation surface whenever there is no infalling matter (and gravitational waves)~\cite{EngWal18}; this matter may in principle be removed by acting with simple sources~\cite{EngWal18, EngPen21}, however quantum backreaction introduces a number of subtleties discussed in~\cite{EngPen21b}.

Returning to the construction at hand, since any two purifications of $\rho_{\rm BH}(t)$ are unitarily related,\footnote{For this to be true, the two Hilbert spaces should have the same
dimension. One might wonder if $\mathcal{H}_{\rm RAD}$ is too large to be unitarily related to $\mathcal{H}_{\rm BH}$. However, this should not cause concern: most of the
Hilbert space $\mathcal{H}_{\rm RAD}$ is not explored by the Hawking radiation -- 
we only permit modes into $\mathcal{H}_{\rm RAD}$ that could fit into $\mathcal{H}_{\rm BH}$ in the
    first place. Thus, we can divide $\mathcal{H}_{\rm RAD} = \mathcal{H}_{\rm ENT} \otimes \mathcal{H}_{\rm AUX}$ with $|\mathcal{H}_{\rm ENT}|=|\mathcal{H}_{\rm BH}|$, and act with a unitary on the radiation to get a state that factorizes on $\mathcal{H}_{\rm ENT} \otimes \mathcal{H}_{\rm AUX}$, so that the factor $\mathcal{H}_{\rm ENT}$ contains all entanglement with the black hole.} 
    there exists a unitary
    $U_{\rm RAD}$ with support only on $\mathcal{H}_{\rm RAD}$ that satisfies
\begin{equation}
\begin{aligned}
    U_{\rm RAD}\otimes \mathbb{I}_{\rm{BH}} \ket{\Psi(t)} = \ket{\sqrt{\rho_{\rm BH}(t)}}.
\end{aligned}
\end{equation}
Here $U_{\rm RAD}$ can be seen as a quantum computation on the radiation that gives a semiclassical geometrization of the entanglement between the radiation and the black hole. 
Effectively, $U_{\rm RAD}$ collapses the Hawking radiation into a black hole and then
shortens the resulting wormhole as much as is possible when only acting on the
radiation. Since $U_{\rm RAD}$ has no
support on $\mathcal{H}_{\rm BH}$, the entanglement wedge of $\rm BH$ is left unchanged. So is the
entanglement spectrum of the black hole.

The procedure works equally well for two-sided black holes. With two CFTs on $\mathcal{H}_{\rm BH_L}\otimes
\mathcal{H}_{\rm BH_R}$, with the system on $\mathcal{H}_{\rm BH_R}$ coupled to $\mathcal{H}_{\rm RAD}$,
$\CPS{\rho_{\rm BH_R}(t)}$  geometrizes the entanglement between $\rm BH_{R}$ and $\rm BH_{L} \cup RAD$.\footnote{Note that in this case we may also choose to canonically purify the state on $\mathcal{H}_{\rm BH_L}\otimes
\mathcal{H}_{\rm BH_{R}}$, resulting in $\ket*{\sqrt{\rho_{\rm BH_{R}\ BH_{L}}(t)}}$. The same line of reasoning applies.}
Now the unitary relating the full state $\ket{\Psi(t)}\in \mathcal{H}_{\rm BH_L}\otimes
\mathcal{H}_{\rm BH_{R}}\otimes \mathcal{H}_{\rm RAD}$ to the canonical purification has support only on
$\mathcal{H}_{\rm BH_L}\otimes \mathcal{H}_{\rm RAD}$ and effectively 
throws the radiation into the left black hole and shortens the wormhole as much as
possible while leaving the right black hole alone.

Let us emphasize a few aspects of our construction. First, the canonical purification of the (right) black hole is not identical to the
thermofield double and can never be made into the TFD as long as we only act on the radiation (and left black hole).
Second, while it might be intuitive that some action on the Hawking radiation, like collapsing it into a black hole, can create a semiclassical geometric connection to the original
black hole, there has to our knowledge not been any explicit demonstration of a unitary implementing this in broad generality.  Our protocol relies crucially on the recently discovered post Page time QES. In fact, more surprisingly, we will show in Sec.~\ref{sec:prePage} that in the absence of such a QES, it is not possible to construct a wormhole connecting an unmodified ${\cal W}_{E}[\rho_{\mathrm{BH}}]$ to the radiation even when the two are highly entangled, showing that in the absence of significant modifications, a strict interpretation of ER=EPR will not hold.

\subsection{A Simple AdS/CFT Dictionary for Unitary Invariants}
Because $\CPS{\rho_{\rm BH}(t)}$ is unitarily related to $\ket{\Psi(t)}$ via a unitary with no support on $\rm BH$, $S_{\rm vN}[\rho_{\rm RAD}(t)]$ can be computed in the CPT conjugated spacetime. In particular, computing $S_{\rm vN}[\rho_{\rm RAD}(t)]$ in $\ket*{\sqrt{\rho_{\rm BH}(t)}}$ requires no modification of the homology constraint of the QES: $\ket*{\sqrt{\rho_{\rm BH}(t)}}$ is a standard instance of AdS/CFT.  The
entropy of the complement of $\rm BH$ is computed by $S_{\rm gen}[\chi[t]]$, and
 the state on the complement of $\rm BH$ is unitarily related to $\rho_{\rm RAD}(t)$; thus $S_{\rm
gen}[\chi[t]] = S_{\rm vN}[\rho_{\rm RAD}(t)]$. More generally, any quantity that is invariant under local unitaries will be identical in the two states, giving a way of computing unitary invariants without novel modifications.\footnote{It would be interesting to understand the import on the work of~\cite{GenKar21} regarding implications of relaxing the homology constraint.}

\begin{figure}
     \centering
     \includegraphics[width=1\textwidth]{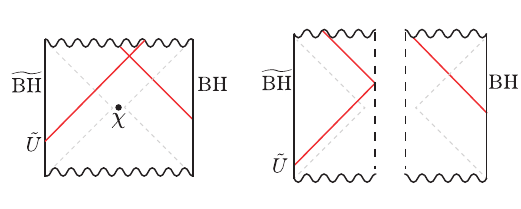}
     \caption{Canonical purifications after (left) and before (right) the Page time. 
     After the Page time, a signal sent from $\widetilde{\rm BH}$ via some local unitary $\tilde{U}$ can effect an infalling observer from $\rm  BH$. Before the Page time, this cannot happen.}
     \label{fig:infalling}
\end{figure} 

The standard AdS/CFT picture of the radiation presented by $\CPS{\rho_{\mathrm{BH}}}$ provides a simple geometrization of the difference in the experience of an infalling observer before and after the Page time. If we turn on a sufficiently early local unitary in $\widetilde{\rho_{\mathrm{BH}}}[t>t_{P}]$, an infalling observer will encounter the energy shock resulting from this unitary. This is illustrated in Fig.~\ref{fig:infalling}. We naturally expect that acting on the radiation can modify the experience of the infalling observer after the Page time; the dual of $\CPS{\rho_{\mathrm{BH}}}$ precisely geometrizes this expectation, yielding the same setup as in Marolf-Wall~\cite{MarWal12}. Indeed, it is geometrically clear that there is no localized operator acting on $\widetilde{\rho_{\mathrm{BH}}}[t<t_{P}]$ that can modify the experience of an infalling observer prior to the Page time. See Fig.~\ref{fig:infalling}.

\subsection{Concrete Example: JT Gravity with Conformal Matter}
To alleviate any concerns about potential sick behavior of the canonical purification, let us now construct fully explicit canonical purifications of the evaporating
one-sided black hole in JT gravity coupled to conformal matter. Indeed, we will see that the average null energy condition holds and that the QES is acausally separated from the conformal boundary. We first summarize the salient results from~\cite{EngMer16, AEMM}.

\subsubsection*{The Evaporating Black Hole in JT Gravity}
We work with JT gravity \cite{Tei83, Jac85, AlmPol14} coupled to a CFT
\begin{equation}
\begin{aligned}
    I_{\rm JT} &= I_0[g_{ab}] + I_{g}[g_{ab}, \phi] + I_{\rm CFT}[g_{ab}],
    \\
        I_0 &= \frac{  \phi_0 }{ 16 \pi G }\left[\int_M R  + 2 \int_{\partial M}K
        \right], \\
        I_{g} &= \frac{ 1 }{ 16\pi G }\left[ \int_{M} \phi (R+2) +
        2\int_{\partial M}\phi (K-1) \right], \\
\end{aligned}
\end{equation}
which on-shell yields a locally AdS$_{2}$ metric  with the following dilaton equations of motion:
\begin{equation}
\begin{aligned}
-\nabla_a \nabla_b \phi + g_{ab}\nabla^2 \phi - g_{ab}\phi = 8\pi G_N T_{ab},
\end{aligned}
\end{equation}
where $T_{ab}$ is the CFT stress tensor. We work in the semiclassical approximation where we replace $T_{ab}$ with $\left<T_{ab}\right>$, and omit brackets from now on.

We take the boundary $\partial M$ of the AdS$_2$ spacetime to lie at a finite cutoff, with boundary
conditions
\begin{equation}\label{eq:BC}
\begin{aligned}
    h_{uu} = \frac{ 1 }{ \epsilon^2 }, \qquad \phi|_{\partial M} = \frac{
        \bar{\phi}_r }{ \epsilon },
\end{aligned}
\end{equation}
for constants $\bar{\phi}_r$ and $\epsilon\ll 1$, where $h_{uu}$ is the induced
metric on the boundary, and $u$ the boundary time. We will make use of Poincare coordinates,
\begin{equation}
\begin{aligned}
\dd s^2 = -\frac{ 4 \dd x^+ \dd x^-}{ (x^+-x^-)^2 } = \frac{ -\dd t^2 + \dd z^2 }{
    z^2}, \qquad x^{\pm} = t\pm z,
\end{aligned}
\end{equation}
and describe the boundary location by the function $f(u)$, which gives its
Poincare time as function of boundary time: $t|_{\partial M} = f(u)$.
Solving \eqref{eq:BC} to leading order in $\epsilon$ then gives $z|_{\partial M}
= \epsilon f'(u)$.

A two-sided static JT black hole at temperature $T$ and with energy $E$ is given by
\begin{equation}\label{eq:phiBH}
\begin{aligned}
    \phi = 2\bar{\phi}_r \frac{ 1 - (\pi T)^2 x^+ x^- }{ x^+-x^- }, \qquad E=
    \frac{ \pi \bar{\phi}_r }{ 4G_N }T^2.
\end{aligned}
\end{equation}
In \cite{AEMM} a black hole at temperature $T_0$ was
coupled to a bath CFT$_2$ living on a half-line by turning on absorbing boundary
conditions at the AdS boundary for $t\geq 0$. Instantly turning on absorbing boundary conditions leads to the injection
of a positive energy shockwave into the bulk that raises the temperature of the black hole to $T_1
> T_0$ at $t=0^+$, and after that the effective temperature falls off as the
black hole evaporates.  
The resulting bulk stress tensor is
\begin{equation}\label{eq:TAEMM}
\begin{aligned}
    T_{x^- x^-}(x^-)  = E_S \delta(x^-) - \frac{ c }{ 24\pi }\{
        f^{-1}(x^-), x^-\} \theta(x^-), \qquad T_{x^+x^+}=T_{x^+x^-} = 0,
\end{aligned}
\end{equation}
where $E_S$ is the injected shockwave energy, $c$ the central charge of the bulk CFT,
$\{\cdot, \cdot\}$ the Schwarzian derivative, and $\theta$ a step function.
After determining the stress tensor, the backreaction on the
dilaton can be found using the fact that the dilaton equations of motion can be
directly integrated when $T_{x^+x^-}=0$ \cite{AlmPol14}.

Now, to explicitly solve for the boundary trajectory $f(u)$, it is useful to have the spacetime energy $E(u)$, which is given by:
\begin{equation}\label{eq:Esol}
\begin{aligned}
    E(u) &= \theta(-u) E_0 + \theta(u)E_1 e^{-ku},  \qquad k = \frac{ c G_N }{ 3 \bar{\phi}_r },
\end{aligned}
\end{equation}
where $E_0 = \frac{ \pi\bar{\phi}_r }{ 4G_N }T_0^2$ and $E_1 = E_0 + E_S = \frac{ \pi \bar{\phi}_r }{ 4G_N }T_1^2$. \eqref{eq:Esol} together with \eqref{eq:Eschw} in the appendix can be used to solve for $f(u)$. 

\subsubsection*{The Construction}
Now let us turn to the case at hand: the explicit construction of the canonical purification after the Page time, 
which here means $u \sim \mathcal{O}(k^{-1})$.
We can either consider the same setup as in \cite{AEMM}, meaning we work with a
two-sided black hole evaporating on one side, or we can replace the left conformal boundary with an (unflavored)
end-of-the-world brane, so that our spacetime is one-sided, as in~\cite{KouMal17}. 
The effect of this modification is that the early time QES becomes the empty
surface, which is now homologous to the right conformal boundary. The late time
QES is unchanged, giving thus the same late time canonical purification.\footnote{We assume the brane has sufficient tension so
that, at $t=0$, the location of the brane approaches that of the would-be physical conformal boundary defined by
\eqref{eq:BC}. }

To construct the canonical purification at some fixed boundary time $u_{\partial}$, 
we do the following: (1) take a spatial slice
$\Sigma$ anchored at the late time QES and the boundary at $t_{\partial}=f(u_{\partial})$ and compute the dilaton
and CFT stress tensor on $\Sigma$, (2)
glue $\Sigma$ to its CPT conjugate slice $\tilde{\Sigma}$ across the QES, and 
(3) evolve the new initial data $\Sigma \cup \tilde{\Sigma}$ with some choice of boundary conditions.

Any choice of boundary conditions will give the same spacetime in the Wheeler-de-Witt patch $D[\Sigma\cup \tilde{\Sigma}]$, and so the geometry 
encoded in the canonical purification at time $t_{\partial}$, which encodes the
entanglement structure between the black hole and the Hawking radiation at that time, does not care about boundary
conditions. However, it is interesting to see a more complete development of the spacetime, and so we will choose to evolve 
the state beyond $D[\Sigma\cup \tilde{\Sigma}]$ with  reflecting boundary conditions.
The latter means that there is no flux of energy across the conformal boundary:
\begin{equation}\label{eq:refcond}
\begin{aligned}
T_{x^-x^-}|_{\partial M} - T_{x^+x^+}|_{\partial M} =0.
\end{aligned}
\end{equation}
This seemingly presents a problem: the data 
on $\Sigma$ is just the data inherited from the evaporating black hole, so 
for any $u_{\partial}>0$ and choice of $\Sigma$, the left hand side of \eqref{eq:refcond} is
nonzero. One way to deal with this is to turn off the transparent boundary
conditions over a small time window $2 \delta u$ around $u_{\partial}$, which leads to the injection of
an energy shock and imperfect reflectivity over the time window $[u_{\partial} - \delta u, u_{\partial} + \delta u]$. 
We can work in the limit $\delta u \rightarrow 0$ so that the injected shock becomes a delta function.
In this case, evolving $T_{ab}$ with reflecting boundary conditions translates into finding a distributional solution of
$\nabla_a T^{ab}=0$ that agrees with the stress tensor in $D[\Sigma]\cup D[\tilde{\Sigma}]$, 
and which satisfies reflecting boundary conditions in the distributional sense:
\begin{equation}
\begin{aligned}
    \int_{u_0 - \delta u}^{u_0 + \delta u}\dd u \left[ T_{x^-x^-}(u) - T_{x^+x^+}(u)\right]|_{\partial M} = 0, \qquad
    \forall u_0 \in \mathbb{R}, \delta u>0.
\end{aligned}
\end{equation}
 
\subsubsection*{Initial data}
Let now $u_{\partial}\sim \mathcal{O}(k^{-1})$ be after the Page time. 
We begin by finding the initial data to be evolved in Poincare coordinates. After this we 
transition to global coordinates, where we do the full evolution.

The late time QES and the boundary time in question is strictly to the future of the 
shockwave arising from turning on the coupling to the bath, so the stress tensor is just given by the Schwarzian term in \eqref{eq:TAEMM}. As shown in \cite{AEMM}, for $u\sim\mathcal{O}(k^{-1})$, we have
\begin{align}
    \{f^{-1}(u), u \} &= \frac{ 1 }{ 2(u-t_{\infty})^2 }\left(1 + \mathcal{O}(k^2e^{-ku})\right), \label{eq:lateschwarz}
\end{align}
where $t_{\infty}$ is the Poincare time at which the dilaton boundary terminates to the future in the original evaporating black hole spacetime. 
It can be checked that the approximation \eqref{eq:lateschwarz} is valid for all $x^-$ covered by a spatial
slice running from the boundary to the QES, and so our stress tensor initial data on the interior of $\Sigma$, or more
precisely in $D[\Sigma]$, reads
\begin{equation}\label{eq:Tmmcont}
\begin{aligned}
    T_{x^- x^-}|_{D[\Sigma]}\approx -\frac{ c }{ 48\pi } \frac{ 1 }{ (x^- - t_{\infty})^2 }.
\end{aligned}
\end{equation}
As we alluded to earlier, additional shocks will be present at $\partial \Sigma$ in the canonical purification, i.e.\ at the QES and at the conformal boundary. We will return to these shocks in a moment.

The calculation of the dilaton in $D[\Sigma]$ is somewhat more involved, and carried out in the appendix. The final result is
\begin{equation}\label{eq:phiDataLate}
\begin{aligned}
    \phi(x^+, x^-)|_{D[\Sigma]} &= \frac{\bar{\phi}_r}{2(x^+ - x^-)} \Big[(2 t_\infty - x^- -
    x^+)\mathcal{C}
    -k(x^+ - x^-) \\
    &\qquad \qquad \qquad \qquad + k(2 t_{\infty} - x^+ - x^-)\log\left( \frac{ t_{\infty}-x^- }{ t_{\infty} }\right)
    \Big],
\end{aligned}
\end{equation}
where we have neglected terms of order $\mathcal{O}\left([t_{\infty}-x^{\pm}]^2\right)$ in the square brackets,
which are suppressed since $t_{\infty} - x^{\pm}$ is non-perturbatively small on the whole of $\Sigma$.
The constant $\mathcal{C}$ is
\begin{equation}
\begin{aligned}
    \mathcal{C} = \frac{ 4 }{ t_{\infty} }+k(2+\gamma),
\end{aligned}
\end{equation}
where the $\mathcal{O}(1)$ constant $\gamma$ is defined in the appendix.

\subsubsection*{Changing coordinates}
We have thus far determined the initial data on $D[\Sigma]$. Let us now transition to
global coordinates, where we will work out the shocks and evolve our initial data. We define global coordinates through
\begin{equation}\label{eq:globalcoords}
\begin{aligned}
    \mu(t+\lambda) &= \frac{ \sin(\tau-\tau_0) }{ \cos(\tau-\tau_0) + \sin \rho }, \qquad  \rho \in \left(-\frac{ \pi }{ 2 }, \frac{ \pi }{ 2
    }\right), \\
    \mu z &= \frac{ \cos \rho }{ \cos (\tau-\tau_0) + \sin \rho },
\end{aligned}
\end{equation}
where $\lambda, \tau_0 \in \mathbb{R}$ and $\mu >0$ are constants parametrizing isometries that we can choose freely to
obtain the most convenient coordinates. This brings the metric to the form
\begin{equation}
\begin{aligned}
    \dd s^2 = \frac{ 1 }{ \cos^2 \rho }\left(-\dd \tau^2 + \dd \rho^2\right) =
    \frac{ -\dd w^+ \dd w^{-}  }{ \cos^2
    \left(\frac{ w^+ - w^- }{ 2 } \right) }, \qquad w^{\pm} = \tau \mp \rho,
\end{aligned}
\end{equation}
with the right conformal boundary is at $\rho=\frac{ \pi }{ 2 }$.

For convenience we now want to pick $\lambda, \mu, \tau_0$ so that our initial data slice $\Sigma$ can be taken to be the $\tau=0$ slice, 
and with the QES at $\rho=0$. 
The solution is worked out in the appendix. The result is
\begin{align}
    \chi & \equiv \mu(t_{\infty}+\lambda) = -\frac{ 1 }{ 4 } 
    + \frac{ 5 e^{-k u_{\partial}/2 } k }{ 8\pi T_1  } 
    + \mathcal{O}(k^2 ),  \label{eq:chidef} \\
    \tau_0 &= \frac{ \pi }{ 2 } - \frac{ k e^{-k u_{\partial}/2}  }{ \pi T_1 } + \mathcal{O}(k^2),
    \label{eq:tau0sol} \\
    \mu &= \frac{ 1 }{ t_{\infty} - f(u_{\partial})}\left(\frac{ 3 }{ 4 } -
    \frac{ 3k e^{ku_{\partial}/2} }{  8 \pi T_1 } + \mathcal{O}(k^2)
    \right). \label{eq:musol}
\end{align}

\subsubsection*{Determining the shocks}
Next, let us work out the shocks present at the QES caused by the discontinuous derivatives of the dilaton
after gluing $\Sigma$ to $\tilde{\Sigma}$. In the appendix, we derive the null junction conditions in JT gravity and show
that a null junction $S$ with a future-directed generator $k^a$ (unrelated to the constant $k$) induces a stress tensor shock
\begin{equation}
\begin{aligned}
    T_{ab}\ell^a \ell^b = -\frac{ 1 }{ 8\pi G_N }\delta(\lambda)[\ell^a \nabla_{a}\phi],
\end{aligned}
\end{equation}
where $\ell^a = \left(\frac{ \dd }{ \dd \lambda}\right)^a$ is the null tangent of a future-directed (not necessarily
affine) geodesic with $k\cdot\ell = -1$ and $\lambda=0$ on $S$. The bracket denotes the discontinuity across the junction in the direction from past to future.

\begin{figure}
     \centering
     \includegraphics[width=0.5\textwidth]{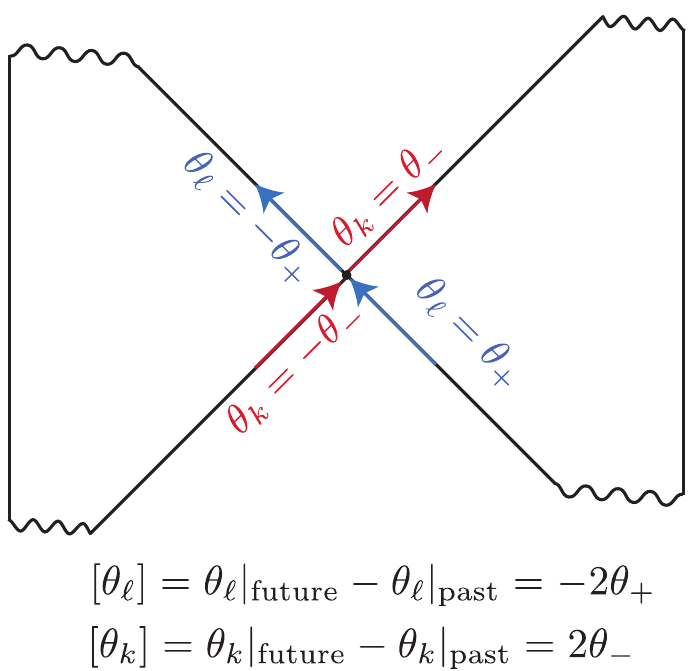}
     \caption{Illustration of the discontinuities of the null expansions across the QES, where $\theta_{\pm}$ are the expansions in the right wedge. }
     \label{fig:junctionDiscont}
\end{figure} 

Consider now a specific set of null-vectors:
\begin{equation}
\begin{aligned}
    \tilde{\ell}^{\mu} = (\partial_{w^+})^{\mu}, \qquad
    \tilde{k}^{\mu} = 2 \cos\left(\frac{ w^+ - w^- }{ 2 } \right)^2(\partial_{w^-})^{\mu},
\end{aligned}
\end{equation}
and let us focus on the junction generated by $\tilde{k}^a$. See Fig.~\ref{fig:junctionDiscont}. For this normalization, $\lambda = w^+$. Defining now the expansion of the QES as we approach it from
$D[\Sigma]$ to be
\begin{equation}
\begin{aligned}
    \tilde{\theta}_+ \equiv \tilde{\ell}^a\nabla_a \phi|_{\text{QES}, \Sigma } = \partial_{w^+}\phi|_{\text{QES}, \Sigma },
\end{aligned}
\end{equation}
we have
\begin{equation}
\begin{aligned}\label{eq:Tppshock}
    T_{w^+w^+}|_{\rm shock} = - \frac{ 1 }{ 8\pi G_N
    }\delta(w^+)[\theta_{\tilde{\ell}}] = \frac{ \tilde{\theta}_+ }{ 4\pi G_N
    }\delta(w^+) \equiv a_+ \delta(w^+).
\end{aligned}
\end{equation}
The sign of $[\theta_{\tilde{\ell}}]$ is explained in Fig.~\ref{fig:junctionDiscont}.
A completely analogous computation gives that
\begin{equation}\label{eq:Tmmshock}
\begin{aligned}
    T_{w^- w^-}|_{\text{QES shock}} = -\frac{  \tilde{\theta}_- }{ 4\pi G_N
    }\delta(w^-) \equiv a_- \delta(w^-), \qquad
    \tilde{\theta}_- = \partial_{w^-} \phi|_{\text{QES}, \Sigma }.
\end{aligned}
\end{equation}
Again the sign is explained in Fig.~\ref{fig:junctionDiscont}. Computing the dilaton derivatives at
$w^+=w^-=0$, we find to leading order
\begin{equation}
\begin{aligned}
    \label{eq:aplusmin}
    a_+ &= - \frac{ c }{ 12\pi } + \mathcal{O}(k\log k),  \\
    a_- &= \frac{ c }{ 12\pi } \frac{ e^{-u_{\partial}k/2} }{ t_{\infty} k } +
    \mathcal{O}(\log k).
\end{aligned}
\end{equation}
We see that $T_{w^+w^+}$ has negative null energy, but it is easy to check that the ANEC holds for any $b>0$.

\subsubsection*{Evolving the stress-tensor}
We have now obtained a convenient coordinate system where the QES lies at $(\tau, \rho)=(w^+, w^-)=(0,0)$,
and where $\Sigma$ can be taken to be the slice $(\tau=0, \rho\geq 0)$.
Let us now change coordinates to prepare for
evolution of the data. Combining \eqref{eq:jacobian}, \eqref{eq:Tmmcont}, and \eqref{eq:chidef}, we 
get
\begin{equation}\label{eq:gdef}
\begin{aligned}
    g(w^- \geq 0)& \equiv T_{w^-w^-}|_{D[\Sigma]} = \left(\frac{ \partial w^- }{ \partial x^- }\right)^{-2}
    T_{x^-x^-}|_{D[\Sigma]}  \\
     &= - \frac{ c }{ 48\pi }\frac{ 1 }{ \left[\chi+ \chi \sin(w^--\tau_0) + \cos(w^--\tau_0)\right]^2}.   \\
\end{aligned}
\end{equation}

We can now finally write down the initial data on the full Cauchy slice:\footnote{
Of course, it is natural to take $\Sigma$ to be the $\tau=0$ slice, in which case we set $-w^+=w^-=\rho$ above, 
but it is convenient to work directly with $w^{\pm}$ coordinates. }
CPT conjugation acts simply on our new coordinates,
\begin{equation}
\begin{aligned}
    \text{CPT}(w^+, w^-) = (-w^+, -w^-),
\end{aligned}
\end{equation}
and so to glue $\Sigma$ to its CPT conjugate, all we need to do for the stress tensor is to extend it symmetrically
about $\rho=0$. For the function $g(w^-)$, we must extend $g$ symmetrically about $w^-=0$.
In toto we have
\begin{equation}\label{eq:initialdataT}
\begin{aligned}
    T_{w^- w^-}|_{\Sigma \cup \tilde{\Sigma}} &= g(w^-) + a_- \delta(w^-),
    \\
    T_{w^+ w^+}|_{\Sigma \cup \tilde{\Sigma}} &= a_+ \delta(w^-).
\end{aligned}
\end{equation}
As explained earlier, we now want a distributional evolution that (1) agrees with \eqref{eq:initialdataT}
on $D[\Sigma]\cup D[\tilde{\Sigma}]$ and (2) has reflecting boundary conditions in the distributional
sense.
Finding the solution is straight forward and carried out in the appendix. 
Stress tensor conservation $\nabla_{a}T^{ab}=0$ together with conformality ($T
\indices{^a_a} = 0$) is enough to completely solve for $T_{ab}$ given some initial data.
The solution is simply given by
\begin{equation}
\begin{aligned}
    T_{w^+w^+}=T_{w^+w^+}(w^+), \qquad T_{w^+w^+}=T_{w^-w^-}(w^-),\qquad T_{w^+w^-} =0.
\end{aligned}
\end{equation}
The only complication is working out how reflecting boundary conditions are implemented.
The result on the domain $|w^{\pm}| \leq \pi$ is given in the appendix.

\begin{equation}\label{eq:Tevolution}
\begin{aligned}
    T_{w^-w^-}(w^-) &= 
    b\delta\left(w^{-}-\frac{ \pi }{ 2 }\right) + b\delta\left(w^{-}+\frac{ \pi }{ 2 }\right) + a_{+}\delta(w^-+\pi)
    + a_+ \delta(w^--\pi) + a_- \delta(w^-), \\
    &\quad + g(w^-) \theta\left(w^- - \frac{ \pi }{ 2 }\right)\theta\left(\frac{ \pi }{ 2
    }-w^-\right) \\
    T_{w^+w^+}(w^+) &= 
    b\delta\left(w^{+}-\frac{ \pi }{ 2 }\right) + b\delta\left(w^{+}+\frac{ \pi }{ 2 }\right) + a_{-}\delta(w^+ + \pi) +
    a_- \delta(w^+ - \pi)  +  a_+ \delta(w^+),  \\
    &\quad +\theta\left(w^+ - \frac{ \pi }{ 2 }\right)g(w^+ - \pi) + \theta\left(-\frac{ \pi }{ 2 }-w^+\right)g(w^+ +
    \pi),
\end{aligned}
\end{equation} 
The solution depends on the constant $b$ that gives the injected energy shock caused by turning off absorbing boundary conditions. Its value depends on the CFT dynamics and how the coupling is turned off. However, from previous studies \cite{AEMM, EngMer16} we know that $b>0$.

\begin{figure}
     \centering
     \includegraphics[width=0.6\textwidth]{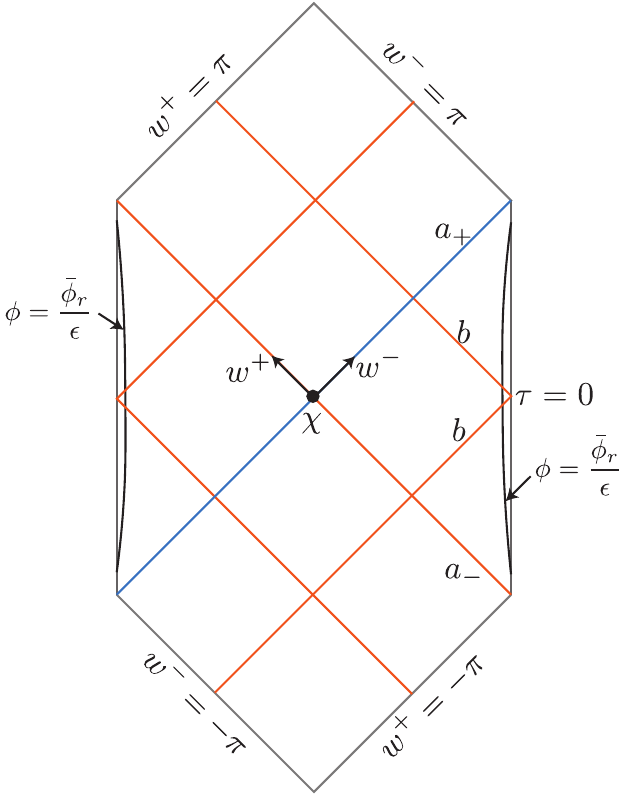}
     \caption{Canonical purification of the evaporating black hole in JT gravity. The orange lines show positive null energy shockwaves, while the blue line is a negative null energy shock. }
     \label{fig:JT_CP}
\end{figure} 

We illustrate the shock profiles in Fig.~\ref{fig:JT_CP}. Note that we do not care about $|w^{\pm}|> \pi$ for the following reason: 
we only need to keep track of the spacetime that is dual to the physical conformal boundary, which is the union of all
the WdW patches of the physical conformal boundary. Now, if the QES is to be
acausally separated from the boundary, for which $b>0$ is sufficient (see the appendix),
then the physical conformal boundary must terminate to the future and past at some $|\tau| \leq \frac{ \pi }{ 2 }$. 
In null coordinates this translates to the fact that the part of spacetime we are interested in is contained in
the region $|w^{\pm}|\leq \pi$.

\subsubsection*{Evolving the dilaton}
Changing to global coordinates, the general solution for
the dilaton \eqref{eq:phisol} is 
\begin{equation}\label{eq:generalglobal}
\begin{aligned}
    \phi(w^+, w^-) &= \frac{ c_1 + c_2(f_+ + f_-) + c_3 f_+ f_- + 8\pi G_N(\hat{I}_+ + \hat{I}_-) }{ f_+ - f_- } \\
    \hat{I}_+(w^+, w^-) &= \int_{-\varepsilon}^{w^+} \dd s \left[1-\sin(s-\tau_0)\right]\left[f_+(s) - f_+ \right]\left[f_+(s) - f_-
    \right]T_{w^+w^+}(s), \\
    \hat{I}_-(w^+, w^-) &= -\int_{\varepsilon}^{w^-} \dd s \left[1+\sin(s-\tau_0)\right]\left[f_-(s) - f_+ \right]\left[f_-(s) - f_-
    \right]T_{w^-w^-}(s), \\
\end{aligned}
\end{equation}
where implicit arguments in $f_{\pm}$ always mean $f_+ \equiv f_+(w^+)$, $f_- \equiv f_-(w^-)$. 
We have here chosen the reference point  $(w^+,w^-)=(-\varepsilon, \varepsilon)$ in the integrals to lie 
slightly to the right of the QES inside $D[\Sigma]$, with $\varepsilon>0$ arbitrarily small.

Getting the final value of the dilaton is now a matter of (1) changing to global coordinates in the dilaton initial data \eqref{eq:phiDataLate},
(2) matching onto \eqref{eq:generalglobal} to extract the coefficients $c_i$, and (3) using \eqref{eq:Tevolution} 
to compute the integrals $\hat{I}_{+}$ and $\hat{I}_{-}$. The integrals split up into $\delta$-function contributions and contributions from the continuous part of the stress tensor. The $\delta$-function
contributions effectively add jumps to the coefficients $c_{i}$ as we cross shocks, so we redefine $c_i$ to contain these steps.
Obtaining the final dilaton is straightforward but somewhat involved, and so is carried out in the appendix. The final
result for $w^- > 0 $ reads
\begin{equation}
\begin{aligned}
    \phi(w^+, w^-) &= \phi_{0}(w^+, w^-) + \phi_{m}(w^+, w^-), \\
    \phi_0(w^+, w^-) &= \frac{ c_1 + c_2(f_+ + f_-) + c_3 f_+ f_- }{ f_+ - f_- }, \\
    \phi_m(w^+, w^-) &= \frac{ 1 }{ f_+ - f_- }\Big[ 
    H_1(w^+, w^-)\theta\left(\frac{ \pi }{ 2 } - w^-\right) +
    H_2\left(w^+, w^-\right)\theta\left( w^- - \frac{ \pi }{ 2 }\right) \\
    &\qquad \qquad +\theta\left(-w^+ - \frac{ \pi }{ 2 }\right)H_3(w^+, w^-)
    +\theta\left(w^+ - \frac{ \pi }{ 2 }\right)H_4(w^+, w^-) \Big],
\end{aligned}
\end{equation}
where $\phi_0$ is a homogenous solution (away from the shocks), while $\phi_m$ describes the deviation from a static black
hole solution. The piecewise constant coefficients $c_i$ and the functions $H_{i}$ are given in the appendix. 

\section{Puzzles Before the Page Time}\label{sec:prePage}

Let us now examine the canonical purification of a young single-sided AdS black hole evaporating into a reservoir; to begin with, we will focus on single-sided black holes (we will consider multiple boundaries in Sec.~\ref{sec:prePage2}). Before the Page time, the dominant QES defining
the entanglement wedge of $\rho_{\mathrm{BH}}$ is simply $\varnothing$: Cauchy slices of $W_{E}[\rho_{\mathrm{BH}}]$ are inextendible. As discussed in the introduction, the inextendibility of $W_{E}[\rho_{\mathrm{BH}}]$ has two important consequences: first, if the pre-Page Hawking radiation is collapsed into a black hole,  it will not be connected to the original black hole via a semiclassical ERB of the same dimensionality as the original black hole.\footnote{This does not exclude the possibility that collapsing the Hawking radiation into a black hole \textit{and} acting on the original black hole in some nontrivial way could create an ERB.} This follows immediately from inextendibility of Cauchy slices of the pre-Page entanglement wedge ${\cal W}_{E}[\rho_{\mathrm{BH}}]$. Second, a sufficient
amount of bipartite entanglement cannot imply bulk connectedness of a bipartite state\footnote{Let us emphasize this point: while it is clear that a tripartite state will likely require some tripartite entanglement measure, we work with \textit{bipartite} states, and we  still find that the von Neumann entropy falls short.} even under existence of a semiclassical dual bulk satisfying the standard QES formula. These two points show that the strictest interpretation of ER=EPR that does not allow modifications of $\rho_{\mathrm{BH}}$ cannot be correct: we must allow for factorized unitaries. In Sec.~\ref{sec:discussion}, we will discuss the possibility that multipartite entanglement may also play a role, even for bipartite states. 

\subsection{Entanglement Entropy is Not Enough}
Let us remind the reader of the argument for point (1): that the amount of bipartite entanglement is not a sufficient criterion for the emergence of spacetime between bipartite states with holographic semiclassical duals. As discussed in the introduction (since the dominant QES is $\varnothing$), $\ket{\sqrt{\rho_{\mathrm{BH}}[t<t_{P}]}}$
is dual to a two-boundary geometry, which is just two disconnected copies of the original one-sided geometry. The bulk state is just
$\ket{\sqrt{\rho_{\rm bh}}}$, the canonical purification of the bulk state in $\mathcal{W}_{E}[\rho_{\rm BH}]$.

\begin{figure}
     \centering
     \includegraphics[width=0.6\textwidth]{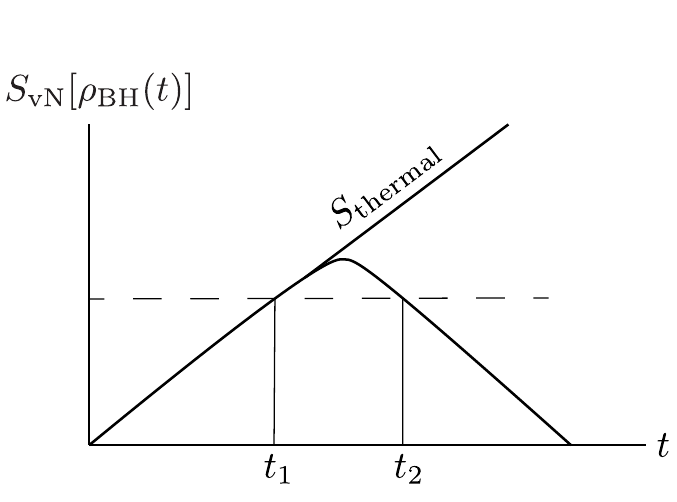}
     \caption{Choice of times to canonically purify.}
     \label{fig:pagecurve}
\end{figure} 

Consider now two boundary times $t_{1}$ and $t_{2}$ such that $t_{1}<t_{\mathrm{P}}<t_{2}$ and
\begin{equation}
S_{\mathrm{vN}}[\rho_{\mathrm{BH}}(t_{1})]= S_{\mathrm{vN}}[\rho_{\mathrm{BH}}(t_{2})].
\end{equation}
This choice is illustrated in Fig.~\ref{fig:pagecurve}. Canonically purifying $\rho_{\mathrm{BH}}(t_{1})$ yields two disconnected
spacetimes whereas canonically purifying $\rho_{\mathrm{BH}}(t_{2})$ yields a single connected spacetime. We may pick $t_{1}$
and $t_{2}$ to be close to $t_{\mathrm{P}}$ so that $S_{\mathrm{vN}}[\rho_{\mathrm{BH}}(t)]$ at both times is large, as in Fig.~\ref{fig:pagecurve}.

When the QES of a connected component of $\mathscr{I}$ is $\varnothing$, it is impossible to add additional bulk regions (connected to $\mathscr{I}$) without modifying the entanglement wedge ${\cal W}_{E}[\rho_\mathscr{I}]$ because Cauchy slices of the entanglement wedge are inextendible. While the formula
\begin{equation}
    S_{\mathrm{gen}} = \frac{\mathrm{Area}[\chi]}{4G} + S_{\mathrm{vN}}[\rho_{\mathrm{out}[\chi]}]
\end{equation}
generally contains an ambiguity in the relative contribution between area and entropy, in the absence of a nontrivial QES, there is no choice of UV cutoff for which there is a surface term. It may in principle be possible to diagnose a nonzero area of a QES purely from the CFT, see~\cite{BelCol21} for an investigation for certain classes of states, although there are a number of subtleties, e.g. \cite{AndPar21}. 

We may take this to be a more general criterion: if the QES of any complete connected component $\mathscr{I}$ is empty, then the bulk dual of the reduced density matrix $\rho_{\mathscr{I}}$ is not semiclassically connected to the remaining spacetime. In particular, under this definition,  a spacetime with just two connected asymptotic boundaries will be disconnected whenever the entanglement wedge of each boundary is inextendible. Note that this is not a necessary condition, as evidenced by two components of two different thermofield double states.

As a quantitative example (and to further illustrate the disconnect between entanglement and spacetime emergence), we will build the canonical purification of a (two-sided) black hole before the Page time in JT gravity in Sec.~\ref{sec:prePage2}. A single-sided example may be constructed by including an (unflavored) end-of-the-world brane, as in~\cite{KouMal17}.

Returning to $\CPS{\rho_{\rm BH}[t<t_{P}]}$ and $\CPS{\rho_{\rm BH}[t>t_{P}]}$, we obtain an immediate counterexample to conjectures claiming that von Neumann entropy necessarily builds
semiclassical spacetime: here, the amount of entanglement is clearly not correlated with spacetime
connectivity. The spacetime can be connected -- as in the post-Page time -- or disconnected -- as before the Page time
-- with the same amount of bipartite entanglement between the two sides. This gives a concrete illustration that bipartite entanglement is not always sufficient for spacetime emergence in a bipartite state, even in standard AdS/CFT. In fact, since our total state is pure, the von Neumann entropy is 
proportional to the reflected entropy and the mutual information, so these quantities cannot diagnose connectivity either. Note that our examples rely crucially on bulk matter entanglement: in the strictly classical case where the HRT
formula applies so that $S_{\rm gen}[\varnothing] =0$, mutual information can always be used to diagnose connectivity. 

A potential complaint at this stage is that while the von Neumann entropy may be identical in the two cases, only the
old black hole is maximally mixed (or more correctly, ``thermally mixed''). We may thus speculate that connectedness
requires our state to have a particular entanglement structure -- for example, the black hole $\rho_{\mathrm{BH}}$ must be  maximally mixed
\textit{in addition} to having a large amount of entropy. In other words, $S_{\mathrm{vN}}[\rho_{\mathrm BH}]$ should be large and $S_{\rm vN}[\rho_{\beta}]- S_{\rm vN}[\rho_{\rm BH}(t)]$ should be
small, where $\rho_{\beta}$ is the thermal state at the same energy and charges as $\rho_{\rm BH}$.  
This modified proposal would then identify connected geometries with states that are (approximately) maximally (thermally) mixed. Prima facie, this appears to diagnose the difference in the canonical purifications before and after the Page time.

However, this turns out to be a red herring. In our example we can easily make $S_{\rm vN}[\rho_{\beta}]- S_{\rm vN}[\rho_{\rm BH}(t)]$ large after the Page time
without altering connectedness. By throwing in a large mass excitation in a pure state from the conformal boundary (this effect can be magnified e.g. by adding a large number of bulk fields), we
can increase $S_{\rm vN}[\rho_{\beta}]$, but since this operation can be implemented by acting with a unitary with
support only on $\mathcal{H}_{\rm BH}$, $S_{\rm vN}[\rho_{\rm BH}(t)]$ is unchanged. One may worry that this operation can modify the QES in some way that alters the connectivity. This concern, however, is unfounded: since $S_{\rm vN}[\rho_{\rm BH}(t)]$ and $S_{\rm gen}[\varnothing]$ are unchanged by the addition of a pure state, we know that even after the additional excitation, 
\begin{equation}
  S_{\rm vN}[\rho_{\rm BH}'(t)]<  S_{\rm gen}[\varnothing]
\end{equation}
where $\rho_{\rm BH}'(t)$ is the state after the addition of the pure state. This immediately implies that even if the dominant QES shifts as a result of the additional matter (and the large number of bulk fields), it remains a nontrivial compact surface and most importantly does not revert to the empty set. Thus the connectivity of the canonically purified old black hole is robust against such modifications. Moving the state away from thermality does not alter its connectivity.~\footnote{See~\cite{GoeLam18} for earlier work on holography of partially mixed states in the purely classical limit.}

\subsection{Complexity is Not Enough}
\begin{figure}
     \centering
     \includegraphics[width=0.6\textwidth]{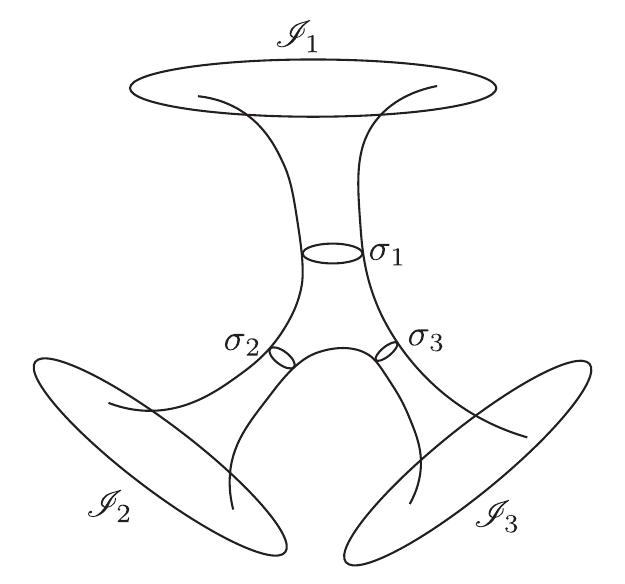}
     \caption{A spatial slice of three-boundary wormhole. Assuming for example that $\sigma_1$ and $\sigma_2\cup \sigma_3$ are QESs with respect to $\mathscr{I}_1$, with $\sigma_2 \cup \sigma_3$ the minimal one, we see that 
     a python's lunch is present, and the region bounded by the $\sigma_i$ is exponentially hard to reconstruct from $\mathscr{I}_1$. }
     \label{fig:wormhole3}
\end{figure}

Another possibility for discerning the feature that is responsible for the disparity in spacetime connectivity between
the two states is computational complexity: it has previously been suggested in work starting with~\cite{Sus14a, Sus14b, Sus14c, StaSus14} that complexity is also
responsible in part of the emergence of the black hole interior. The pre-Page canonical purification is highly complex (more than
one scrambling time after black hole formation), a fact that may be seen geometrically from the existence of a
nonminimal QES in the interior of the entanglement wedge~\cite{BroGha19}. Could it be that exponential complexity is responsible for lack of connection?
The answer appears to be no, as there certainly exist states with exponential complexity (which are also submaximally
mixed), which should by any reasonable definition of connectivity be considered connected. The union of two boundaries of a
three-boundary wormhole, illustrated in Fig.~\ref{fig:wormhole3}, is connected to the third boundary, not maximally mixed with it,
and has non-minimal QESs inherited from the bifurcation surfaces. In the opposite direction, we also expect that at
very early times $\rho_{\mathrm{BH}}(t)$ is not exponentially complex, but $\CPS{\rho_{\mathrm{BH}(t)}}$ is still disconnected. Thus a criterion based on complexity classes fails to diagnose the difference between connected and disconnected spacetimes.

\subsection{An Instructive Counterexample}\label{sec:prePage2}

In light of the failure of bipartite entanglement (and complexity) to build spacetime in $\CPS{\rho_{\mathrm{BH}}(t<t_{P})}$, we must ask what, then, actually builds spacetime. To address this question, we must understand the salient difference between the dual CFT states $\CPS{\rho_{\mathrm{BH}}(t)}$ before and after the Page time.

We will not answer this question in this section (or in this article), but rather provide an additional example that
could serve as an arena to investigate these issues in more detail in the future. 
We will outline the construction of this example in general 
and then compute it explicitly in JT gravity coupled to conformal matter.

For this construction, we work with a two-sided black hole evaporating into a reservoir as in~\cite{AEMM}, 
though we will not restrict ourselves to two dimensions just yet. At any time, we may execute the same procedure as before, tracing out the reservoir and canonically purifying the state $\rho_{\mathrm{BH, LR}}$, which is the joint state of the left and right boundaries. 

\begin{figure}
     \centering
     \begin{subfigure}[b]{0.35\textwidth}
         \centering
         \includegraphics[width=0.9\textwidth]{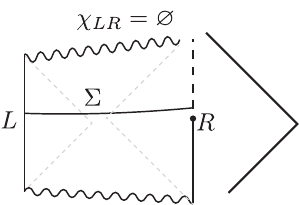}
         \caption{}
     \end{subfigure}
     \begin{subfigure}[b]{0.49\textwidth}
         \centering
         \includegraphics[width=0.9\textwidth]{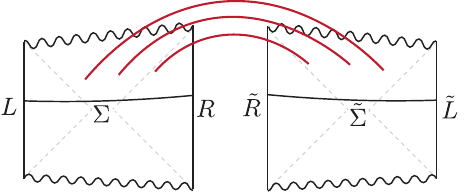}
         \caption{}
     \end{subfigure}
     \begin{subfigure}[b]{0.35\textwidth}
         \centering
         \includegraphics[width=0.9\textwidth]{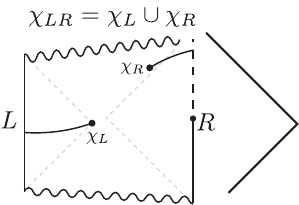}
         \caption{}
     \end{subfigure}
     \begin{subfigure}[b]{0.49\textwidth}
         \centering
         \includegraphics[width=0.9\textwidth]{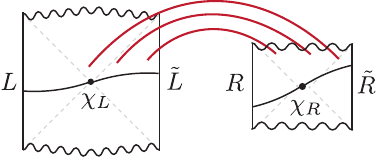}
         \caption{}
     \end{subfigure}
         \caption{In (a) we see an evaporating two-sided black hole with a Cauchy slice $\Sigma$ anchored at $t<t_P$, and in (b) we show the spacetime dual to  $\CPS{\rho_{\rm BH, LR}(t)}$. (c) and (d) shows the same for $t>t_P$. Red lines indicate entanglement of the bulk matter fields.}
        \label{fig:foursided}
\end{figure}

Prior to the Page time, the QES of $\rho_{\mathrm{BH, LR}}(t)$ is the empty set. Canonical purification of the state generates a new two-boundary spacetime \textit{not connected} to the original spacetime via a semiclassical wormhole. This spacetime is illustrated in Fig.~\ref{fig:foursided}a.

Let us now examine this four-boundary, two component spacetime. Label the two original boundaries $L$ and $R$, and the
newly generated ones $\widetilde{L}$ and $\widetilde{R}$. Before the Page time, $L$ is connected to $R$, and $\tilde{L}$ to $\tilde{R}$.
After the Page time however, we have that $L$ is connected to $\tilde{L}$, and $R$ to $\tilde{R}$.
It would be interesting to study various forms of multipartite entanglement between $R, \tilde{R}, L, \tilde{L}$ to
see if different properties of multipartite entanglement may be at play here;\footnote{A similar setup in SYK and JT gravity was studied in \cite{Num20}.}  what amount and type of entanglement
builds spacetime, or is entanglement in fact a red herring and there is some other property of the state that is
responsible for spacetime connectivity?

Let us now construct this spacetime quantitatively in two dimensions. 
In JT gravity it is straightforward to compute the canonically purified spacetime, and we carry it out in the appendix. 
Carrying out the purification at boundary Poincare time $t_{\partial}=f(u_{\partial})$ for $u_{\partial}\sim
\mathcal{O}(1)$, the stress tensor reads, using the same conventions as in Sec.~\ref{sec:postPage},
\begin{equation}
\begin{aligned}
    T_{x^-x^-} &= \theta(t_{\partial}-x^-)h(x^-) + b\delta(x^- - t_{\partial}),\\
    T_{x^+x^+} &= \theta(t_{\partial}-x^+)h(x^+) + b\delta(x^+ - t_{\partial}),
\end{aligned}
\end{equation}
where 
\begin{equation}
\begin{aligned}\label{eq:Tmmearly}
    h(x^-) &=  E_S \delta(x^-) - \frac{ c }{ 24\pi }\theta(x^-) \frac{ 2(\pi T_1)^2 }{ \left[1 - (\pi T_1
    x^-)^2\right]^2 },
\end{aligned}
\end{equation}
where $b>0$ is a constant depending on the microscopics of the CFT and how the absorbing boundary conditions are turning
off -- just like in Sec.~\ref{sec:postPage}.  The dilaton reads
\begin{equation}
\begin{aligned}
    \phi &= \frac{ c_1 + c_2(x^+ + x^-) + c_3 x^+ x^- }{ x^+ - x^- } + 8\pi G_N\frac{\theta(t_{\partial}-x^-)F(x^-)-\theta(t_{\partial}-x^+)F(x^+) }{ x^+ - x^- },
\end{aligned}
\end{equation}
where the piecewise constant coefficents $c_i$ and the function $F$ are listed in the appendix. 

\section{Discussion}\label{sec:discussion}
We have explicitly constructed (1) a precise procedure that generates a connected spacetime from an old black hole and its radiation without modifying the black hole system -- assuming the validity of the QES formula, and (2) an example of a disconnected semiclassical spacetime whose CFT dual is a highly entangled bipartite state. We have ruled out the possibility that this counterexample could be dismissed by appealing to thermality of the old black hole state or by refining the ER=EPR proposal to refer to other entanglement measures such the reflected entropy. In particular, we have shown that no operation that leaves $\rho_{\mathrm{BH}}$ unchanged can consistently construct spacetime connectivity when the bipartite entanglement is large. 

\paragraph{ER=EPR with bipartite entanglement:} If a bipartite state $\ket{\psi}_{AB}$ has a sufficiently large von Neumann entropy, and ERBs are indeed sourced by bipartite entanglement, then there exist unitaries $U_A$ and $U_B$ with support
strictly on $A$ and $B$, respectively, such that $U_A \otimes U_B \ket{\psi}_{AB}$ has a connected semiclassical dual. In light
of the earlier discussions, if this
were to be true, then there would have to exist a unitary $U_{\rm BH}$ such that the state
\begin{equation}
\begin{aligned}
   \rho_{\mathrm{BH}}'= U_{\rm BH}\rho_{\rm BH}(t<t_{P})U_{\rm BH}^{\dagger}
\end{aligned}
\end{equation}
has a non-empty QES. Then with the action of $U_{\rm RAD}$, a Cauchy slice of the entanglement wedge of BH in the state
$U_{\rm BH} \otimes I_{\rm RAD} \ket{\psi}$ could in principle be embedded in a larger spatial slice forming a wormhole. As a constraint on the unitaries in question, they will need to modify the topology of the dominant QES.

\paragraph{Multipartite entanglement for connectivity:} It is clear that for multipartite states $\ket{\psi}_{\mathscr{I}_{1}\cdots \mathscr{I}_{n}}$ (or e.g. subregions), multipartite entanglement measures are necessarily for a proper understanding of the emergent connectivity of the bulk geometry -- e.g. tripartite entanglement for a three-boundary wormhole. However, our counterexample is a strictly bipartite state, where we would have expected bipartite entanglement to be the definitive determinating quantity. Nevertheless, a key difference between $\rho_{\rm BH}(t<t_{P})$ and $\rho_{\rm BH}(t>t_{P})$ is that the former is heavily correlated with itself while the latter is not. It is in principle possible that there is some measure of these internal correlations of a system -- e.g. multipartite entanglement measures -- that may be responsible for the connectivity of the spacetime when the bipartite entanglement turns out to be too coarse.  This appears to have some similarity with the results of~\cite{AkeRat19}, which showed that duals to the entanglement wedge cross-section require holographic states to have tripartite rather than primarily bipartite entanglement; however, it is not obviously the same phenomenon: the arguments of~\cite{AkeRat19} relied on the reflected entropy and entanglement of purification at ${\cal O}(G_{N}^{-1})$ vs. ${\cal O}(1)$, whereas in our case these quantities are ${\cal O}(G_{N}^{-1})$ in both the connected and disconnected phases. Moreover,~\cite{AkeRat19} relied primarily on subregions in their argument, and it is clear that subregions must have at least some non-bipartite entanglement. In our case the state is truly bipartite on two separate boundaries. 

\begin{figure}
     \centering
     \includegraphics[width=0.6\textwidth]{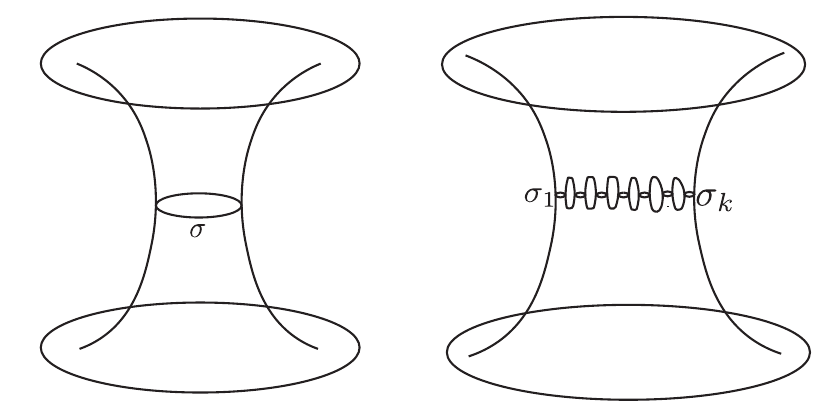}
     \caption{Spatial slices of two different classical two-sided wormholes, both dual to states with the same amount of entanglement between the two sides, i.e.\ $\mathrm{Area}[\sigma] = \mathrm{Area}[\sigma_1 \cup \ldots \cup \sigma_k]$, where $k\sim \mathcal{O}(G_N^{-1})$. The right picture represents a potential toy model for a highly quantum ERB. }
     \label{fig:toymodel}
\end{figure} 

\paragraph{Wormhole condensation:} \cite{AkeEng19a} constructed a toy model for Hawking radiation where the Hawking quanta were
modeled by small classical individual ERBs connected to the original black hole. This models a highly quantum wormhole as a very small ERB. The model is studied at discrete timesteps between the emission of Hawking
quanta, remaining agnostic about the topology-changing process of ERB creation. It would be interesting to see if this
type of toy model also could shed light on the type of quantities that can differentiate between the ``quantum'' and classical wormholes. For instance, we may consider two different two-sided spacetimes with the same $\mathcal{O}(G_N^{-1})$ amount of entanglement,  where one has an HRT surface with one connected component of area $A \sim \mathcal{O}(1)$, while
the other has an HRT surface with $\sim \mathcal{O}(G_N^{-1})$ connected components, each having an area of order
$\mathcal{O}(G_N)$, but with the total area of all components adding up to $A$. See Fig.~\ref{fig:toymodel}. The latter is a toy model of the entanglement before it is merged into semiclassical form. It would
be interesting to see which observables are sensitive to this
fragmentation of the total area.

\paragraph{Code Subspace Dependence:} 
The fact that the entanglement wedge ${\cal W}_{E}[\rho_{\mathrm{BH}}]$ cannot be extended prior to the Page time shows, as argued above, that there is no way to leave $\rho_{\mathrm{BH}}$ unaltered while acting purely on the radiation to create a semiclassical wormhole connecting the two. However, let us pause now for a reminder that while  ${\cal W}_{E}[\rho_{\mathrm{BH}}]$ is the dual to $\rho_{\mathrm{BH}}$, reconstruction of the bulk is quantum error correcting~\cite{AlmDon14} and requires the specification of a nontrivial code subspace.  

As shown in~\cite{HayPen18} and further elucidated in~\cite{AkeLei19}, state-independent reconstruction in general covers only a subset of the entanglement wedge when quantum corrections are taken into account.  In particular,~\cite{HayPen18} argued that it is the entanglement wedge of the maximally mixed state in the code subspace that determines the extent of state-independent reconstruction -- the so-called reconstruction wedge~\cite{AkeLei19}.

If we define the black hole system as the reconstruction wedge of $\rho_{\rm BH}$ in a large code subspace consisting of all states with a smooth horizon, we will in fact find that when the black hole is within the adiabatic regime, the wedge is bounded by a nontrivial QES. In this case, the canonical purification of the maximally mixed state of ${\cal H}_{\mathrm{code}}$ is connected. The potential relevance of the choice of code subspace and typicality for ER=EPR had been previously noted in~\cite{Ver20}, which also clarified the arrow of implication from spacetime connectivity to a high amount of entanglement (conversely with our construction, which illustrates a subtlety into the arrow of implication from EPR to ER).

\paragraph{Experience of an infalling observer:} The experience of an observer falling through the black hole after the Page time can be causally affected by turning on some local unitary in the canonical purification. This gives an explicit realization of action on the island via the radiation, and specifically realizes the map between Marolf-Wall~\cite{MarWal12} and the relation between the radiation and the black hole interior. Note that by extension, this is also an explicit mapping from the factorization problem due to replica wormholes in the island picture to the standard factorization problem in AdS/CFT~\cite{MarWal12, Har15, GuiJaf15, HarJaf18} of multiboundary geometries in AdS/CFT. 

In light of these observations, it is tempting to think of the map $U_{\rm RAD}$ from $\rho_{\mathrm{RAD}}$ to $\widetilde{\rho_{\rm BH}}$ as a decoding unitary. The state $\widetilde{\rho_{\rm BH}}$ is simple, and thus $U_{\rm RAD}$ may be thought of as mapping a high complexity state to a simply reconstructible state. However, it is not a particularly useful `decoding' map: the map $U_{\mathrm{RAD}}$ is dependent on which operators we choose to turn on in the island. Nevertheless, it would be interesting to better understand the properties of this map.

\section*{Acknowledgments} It is a pleasure to thank Sebastian Fischetti, Sergio Hern\'andez-Cuenca, Arjun Kar, Juan Maldacena, and Arvin Shahbazi-Moghaddam for comments on an earlier draft of this manuscript. We are also grateful to Jan de Boer, Sebastian Fischetti, Ben Freivogel, Sergio Hern\'andez-Cuenca, Daniel Jafferis, Arjun Kar, Sam Leutheusser, Henry Lin, Hong Liu, Don Marolf, Geoff Penington, Lisa Randall, Arvin Shahbazi-Moghaddam, Shreya Vardhan, Erik Verlinde, and Herman Verlinde for valuable conversations.  This work is supported in part by the MIT department of physics. NE is supported in part by NSF grant no. PHY-2011905, by the U.S. Department of Energy Early Career Award DE-SC0021886, by the U.S. Department of Energy grant DE-SC0020360 (Contract 578218), by the John Templeton Foundation and the Gordon and Betty Moore Foundation via the Black Hole Initiative, and by funds from the MIT department of physics. The work of \AA{}F is also
supported in part by an Aker Scholarship and by NSF grant no. PHY-2011905. This work was initially started at the KITP in the Gravitational Holography workshop, supported in part by the National Science Foundation under Grant No. PHY-1748958. 
\section{Appendix}
\subsection{JT gravity junction conditions}
The JT equations of motion with matter reads $\tilde{G}_{ab}(\phi) = 8\pi G_N T_{ab}$, where
\begin{equation}
\begin{aligned}
    \tilde{G}_{ab} = -\nabla_{a}\nabla_{b}\phi + g_{ab}\nabla^{2}\phi - g_{ab}\phi
\end{aligned}
\end{equation}
is the effective dilaton ``Einstein tensor''. 
Consider now gluing two spacetimes $M^+$ and $M^-$ along a codimension$-1$ null
junction $S$ generated by a null vector $k^a$. Let $M^+$ be to the future of the junction and
$M^-$ to the past, and define the discontinuity of some quantity $A$ across the junction as
\begin{equation}
\begin{aligned}
  \relax  [A] = A_{+}-A_{-}.
\end{aligned}
\end{equation}
Let now $\ell^a = \left(\frac{ \dd }{ \dd \lambda } \right)^a$ be a rigging null field with $k \cdot \ell = -1$ on $S$.
Firing geodesics along $\ell^a$, the parameter $\lambda$ can be viewed to a scalar field on spacetime in a
neighbourhood of the junction, chosen so that $\lambda=0$ on $S$. Using this, we can in a neighbourhood of the junction write
\begin{equation}
\begin{aligned}
    \phi = \theta(\lambda)\phi^+ + \theta(-\lambda) \phi^-.
\end{aligned}
\end{equation}
Assume that $[\phi]=0$. This is the analogue of the first junction condition and is required in order to have a
distributional Einstein tensor. We then find that
\begin{equation}
\begin{aligned}
    \nabla_a \phi &= \theta(\lambda)\nabla_a \phi^+ + \theta(-\lambda) \nabla_a \phi^-,
    \\
    \nabla_b \nabla_a \phi &= \nabla_{(b} \nabla_{a)} \phi = \delta(\lambda) \nabla_{(b} \lambda[\nabla_{a)} \phi] +
    \text{non-singular}.
\end{aligned}
\end{equation}
Thus the effective Einstein tensor gets a singular part
\begin{equation}
\begin{aligned}
    \tilde{G}_{ab}|_{\rm sing} = \delta(\lambda) \left( -\nabla_{(a} \lambda
        [\nabla_{b)}
    \phi] + g_{ab} \nabla^{c}\lambda [\nabla_c \phi] \right).
\end{aligned}
\end{equation}
By definition we have $\dd x^{\mu} = \ell^{\mu} \dd \lambda$. Contracting this with
$k_{\mu}$ we get
\begin{equation}
\begin{aligned}
k_{\mu} \dd x^{\mu} = - \dd \lambda = - \nabla_{\mu} \lambda \dd x^{\mu} \quad \Rightarrow \quad
\nabla_a \lambda = - k_a,
\end{aligned}
\end{equation}
giving that
\begin{equation}
\begin{aligned}
    \tilde{G}_{ab} = \delta(\lambda)\left(k_{(a}[\nabla_{b)}\phi] - g_{ab}[k^c
    \nabla_c \phi] \right),
\end{aligned}
\end{equation}
or
\begin{equation}
\begin{aligned}
    8\pi G_N T_{ab}\ell^a \ell^{b} = -\delta(\lambda)\left[\ell^a\nabla_a
    \phi\right].
\end{aligned}
\end{equation}

\subsection{Computing the Post Page Time Canonical Purification in JT Gravity}
\subsubsection*{Computing the dilaton on $\Sigma$}
We want the dilaton on a late time slice $\Sigma$ running between the QES and the conformal boundary.
The approach to compute this is the same as in \cite{AEMM}, except we have to relax the assumption $|t_{\infty}-x^+|\ll
|t_{\infty}  - x^{-}|$, which holds near the QES but not near the conformal boundary, and thus not on $\Sigma$ in
general. It is however still true that $\frac{ t_{\infty} - x^{\pm} }{ t_{\infty}}$ is non-perturbatively small.
To make the derivation clearer, we retrace many steps from \cite{AEMM}.

The general solution for the dilaton coupled to matter with $T_{x^+x^-}=0$ reads 
\begin{equation}\label{eq:phisol}
\begin{aligned}
    \phi &= \frac{ c_1 + c_2(x^++x^-) + c_3 x^+ x^- + 8\pi G_N (I^+ + I^-)  }{ x^+ - x^- },\\
            I^+ &= \int_{x^+_{0}}^{x^+}\dd s (s-x^+)(s-x^-) T_{x^+x^+}(s), \\
            I^- &= -\int_{x^-_{0}}^{x^-}\dd s (s-x^+)(s-x^-) T_{x^-x^-}(s).
\end{aligned}
\end{equation}

The boundary trajectory $f(u)$, which will be useful below,
can be found through an analysis of the energy of the spacetime, whose value and rate of change is given by
\begin{align}
    E(u) &= - \frac{ \bar{\phi}_r }{ 8\pi G_N }\{ f(u), u\}, \label{eq:Eschw} \\
    \partial_{u}E(u) &= f'(u)^2\left[T_{x^-x^-}(u) - T_{x^+x^+}(u) \right]. \label{eq:dE}
\end{align}
Equation \eqref{eq:dE} together with the stress tensor \eqref{eq:TAEMM} lets us solve for $E(u)$. Next,
this solution (given in \eqref{eq:Esol}) together with \eqref{eq:Eschw} can be used to solve for $f(u)$.

Now let us work out the dilaton. 
First, from \eqref{eq:phiBH}, \eqref{eq:TAEMM} and \eqref{eq:phisol}, we have that the dilaton to the future of the 
shock reads
\begin{equation}
\begin{aligned}
    \frac{x^+- x^-}{2\bar{\phi}_r} \phi
    &= 1 - (\pi T_1)^2 x^+ x^- + \frac{ k }{ 2 }\int_{0}^{x^-}\dd s (s-x^+)(s-x^-) \{f^{-1}(s), s\},
\end{aligned}
\end{equation}
giving
\begin{equation}\label{eq:dPhi}
\begin{aligned}
    \frac{(x^+- x^-)^2}{2\bar{\phi}_r} \partial_{x^{\pm}}\phi &= \mp \left[1-(\pi T_1)^2 (x^{\mp})^2 + \frac{ 1 }{ 2 }k
    \int_{0}^{x^{-}}\dd s(s-x^{\mp})^2 \{f^{-1}(s), s\} \right].
\end{aligned}
\end{equation}
Let us now approach the endpoint of the conformal boundary at fixed $x^-=t_{\infty}$, meaning $x^+ \rightarrow t_{\infty}$.
This endpoint is characterized by $\phi=0$, and so in order for $\partial_{x^+}\phi$ to stay finite there, we must have
\begin{equation}
\begin{aligned}
    I_{\infty} \equiv \int_{0}^{t_{\infty}} \dd s (s-x^{-})^2 \{f^{-1}(s), s\} = \frac{ 2 }{ k }\left[(\pi T_1 t_{\infty})^2
     - 1 \right].
\end{aligned}
\end{equation}

Let us for later convenience note that
\begin{equation}\label{eq:piT}
\begin{aligned}
     \pi T_1 &= \frac{ 1 }{ t_{\infty} }\left[1 + \frac{ k }{ 4 }t_{\infty}\right] + \mathcal{O}(k^2),
\end{aligned}
\end{equation}
giving
\begin{equation}\label{eq:piCombo}
\begin{aligned}
    1-(\pi T_1)^2 (x^{\pm})^2+\frac{ k }{ 2 }I_{\infty} &=
    (\pi T_1)^2 \left[t_{\infty}-(x^{\pm})^2\right]  \\
    &= \frac{ 1 }{ t_{\infty}^2 }\left[1 + t_{\infty}\frac{ k }{ 2
    }\right](t_{\infty}-x^{\pm})2 t_{\infty} + \mathcal{O}\left((t_{\infty}-x^{\pm})^2\right)+\mathcal{O}(k^2) \\
     &= \left[2 +kt_{\infty} \right]\frac{ t_{\infty}-x^{\pm} }{ t_{\infty} } +
     \mathcal{O}\left((t_{\infty}-x^{\pm})^2\right)+\mathcal{O}(k^2).
\end{aligned}
\end{equation}
From now on we implicitly drop terms of order $\mathcal{O}(k^2)$ and $\mathcal{O}\left((t_{\infty}-x^{\pm})^2\right)$.

Let us now first compute $\partial_{+}\phi$. Using \eqref{eq:lateschwarz} we have
\begin{equation}
\begin{aligned}
    \partial_ - \int_{0}^{x^-}\dd s (s- x^{-})^2 \{f^{-1}(s), s\} &= \mathcal{O}(1) + \int_{}^{x^-}\dd s \frac{ x^{-}-s
    }{ (s-t_{\infty})^2 } \\
    &= -\gamma_1  - \log \left(\frac{ t_{\infty} - x^{-} }{ t_{\infty}  } \right)+\mathcal{O}\left(t_{\infty} - x^- \right)
\end{aligned} \\
\end{equation}
where $-\gamma_1 $ is just some unknown constant which we have parametrized by
$\gamma_1$ for later convenience (note that here and in the following, $\mathcal{O}(1)$ means
$\mathcal{O}\left([t_{\infty}-x^{\pm}]^0\right)$).  We then have
\begin{equation}\label{eq:integral1}
\begin{aligned}
    \int_{0}^{x^-}\dd s (s- x^{-})^2 \{f^{-1}(s), s\} = I_{\infty} + \gamma_1(t_{\infty}-x^-) +
    (t_{\infty}-x^-)\log \left(\frac{ t_{\infty} - x^{-} }{ t_{\infty}  }\right) + \mathcal{O}\left(\left[t_{\infty} - x^-
    \right]^2\right).
\end{aligned}
\end{equation}
Inserting \eqref{eq:integral1} and \eqref{eq:piCombo} into \eqref{eq:dPhi} then gives
\begin{equation}
\begin{aligned}
    \partial_{x^+} \phi &\approx - \frac{\bar{\phi}_r}{(x^+-x^-)^2}\left\{\left[4+kt_{\infty}(2+\gamma_1)\right]\frac{ t_{\infty} - x^{-} }{ t_{\infty} }
     + k(t_{\infty}-x^-)\log \left(\frac{ t_{\infty} - x^{-} }{ t_{\infty}  }\right)\right\}.\label{eq:dPhiPlus}
\end{aligned}
\end{equation}

Next we turn to $\partial_{x^-} \phi$. Expanding $\int_0^{x^-}\dd s(s-x^{+})^2\{f^{-1}(s), s\}$ about $x^+ = t_{\infty}$ in
\eqref{eq:dPhi}, we get
\begin{equation}
    \begin{aligned}\label{partialmin}
        \frac{(x^+ - x^-)^2}{2\bar{\phi}_r}   \partial_{-}\phi &=1 - (\pi T_1)^2 (x^{+})^2 + \frac{ 1 }{ 2 }k
        \mathcal{I}
\end{aligned}
\end{equation}
where
\begin{equation}
\begin{aligned}
    \mathcal{I}&=  I_{\infty} + \mathcal{I}_0 + (x^+-t_{\infty})\mathcal{I}_1 + (x^+-t_{\infty})^2\mathcal{I}_{2},\\
    \mathcal{I}_{0} &= -\int_{x^-}^{t_{\infty}}\dd s(s-t_{\infty})^2 \{f^{-1}(s), s \}, \\
    \mathcal{I}_{1} &= 2\int_{0}^{x^-}\dd s (t_{\infty}-s)\{ f^{-1}(s), s\}, \\
    \mathcal{I}_{2} &= \int_{0}^{x^-}\dd s \{ f^{-1}(s), s\}.\\
\end{aligned}
\end{equation}
Using now again the late time Schwarzian, we get the
leading behavior
\begin{equation}
\begin{aligned}
    \mathcal{I}_{0}  
    &= -\frac{ 1 }{ 2 }(t_{\infty} - x^-) \\
    \mathcal{I}_{1} & 
    \sim \mathcal{O}(1)+ \int^{x^-}\dd s
    \frac{ 1 }{ (t_{\infty}-s) } = - \log\left( \frac{t_{\infty} - x^-}{t_{\infty}}\right) +  \mathcal{O}(1) \\
    \mathcal{I}_{2} &
    \sim \mathcal{O}(1) + \int^{x^-}\dd s \frac{ 1 }{ 2(s-t_{\infty})^2 } = \frac{ 1 }{
        2(t_{\infty} - x^-) } + \mathcal{O}(1) \\
\end{aligned}
\end{equation}
We want to keep the $\mathcal{O}(1)$ term in $\mathcal{I}_{1}$, but we do not need the $\mathcal{O}(1)$ term in
$\mathcal{I}_2$ since it contributes as $\mathcal{O}\left([t_{\infty}-x^+]^2\right)$ to $\mathcal{I}$, which is higher
order than what we work at. Let us define
$\gamma_2$ through
\begin{equation}
\begin{aligned}
    \mathcal{I}_{1} &=  - \frac{ 1 }{
        2} \log\left( \frac{t_{\infty} - x^-}{t_{\infty}}\right) - \gamma_2 + \mathcal{O}(t_{\infty}-x^{-}).
\end{aligned}
\end{equation}
Then, assembling everything, we find
\begin{equation}\label{partialplus}
\begin{aligned}
        \frac{  (x^+ -x^-)^2 }{ \bar{\phi}_r }\partial_{-} \phi &= 
        \left[4+kt_{\infty}(2+\gamma_2)\right]\frac{ t_{\infty}-x^+ }{ t_{\infty} } + k \frac{ (x^+ - t_{\infty})^2 }{ 2(t_{\infty}-x^-) } + k
        (t_\infty-x^{+})\log\left(\frac{t_{\infty}-x^{-}}{t_{\infty}}\right) \\
                    &\quad - \frac{ k }{ 2 }(t_\infty - x^{-}) 
\end{aligned}
\end{equation}
Integrating \eqref{partialmin} and \eqref{partialplus} gives
\begin{equation}
\begin{aligned}
    \frac{ \phi }{ \bar{\phi}_r } &= A(x^-) + \frac{ t_{\infty} - x^{-} }{ x^+ - x^- }\left[\mathcal{C}_1 + k \log\left(
    \frac{ t_{\infty}-x^- }{ t_{\infty} }\right) \right] \\
    \frac{ \phi }{ \bar{\phi}_r } &= B(x^+) + \frac{ k }{ 2 }\log\left(\frac{ t_{\infty} - x^{-} }{ t_{\infty} } \right)
    + \frac{ t_{\infty} - x^+ }{  x^{+}-x^{-} }\left[\mathcal{C}_2 + k \log\left( \frac{ t_{\infty}-x^- }{ t_{\infty} }\right)\right]
\end{aligned}
\end{equation}
for some unknown functions $A$, $B$, where we define
\begin{equation}
\begin{aligned}
    \mathcal{C}_{i} \equiv \frac{ 4+ k t_{\infty}(2+\gamma_{i}) }{ t_{\infty} }.
\end{aligned}
\end{equation}
Taking the difference we find
\begin{equation}
\begin{aligned}
    0 &= A(x^-) - B(x^+) + \frac{ k }{ 2 }\log\left(\frac{ t_{\infty} - x^- }{ t_{\infty} } \right)
    + \frac{ (\mathcal{C}_1 - \mathcal{C}_2)t_{\infty} + \mathcal{C}_1x^+ - \mathcal{C}_2 x^- }{ x^+ - x^{-}
    } 
\end{aligned}
\end{equation}
This equation is inconsistent if we cannot break up each term into one depending purely on $x^+$ and one depending only on
$x^-$, and so we must have that $\mathcal{C}_1 = \mathcal{C}_2 \equiv \mathcal{C}$, leading to $\gamma_1 = \gamma_2$. 
Then we get the solution
\begin{equation}
\begin{aligned}
    B(x^+) &= \text{constant} \equiv B \\
    A(x^-) &= B  - \frac{ k }{ 2 }\log\left(\frac{ t_{\infty} - x^- }{ t_{\infty} } \right)
    + \frac{ 4+2k }{ t_{\infty}  }
\end{aligned}
\end{equation}
and so we get 
\begin{equation}
\begin{aligned}
    \phi = \frac{\bar{\phi}_r}{x^+ - x^-} \left[\mathcal{C}(t_{\infty}-x^+)+B(x^+-x^-)
        + \frac{k}{2}(2t_{\infty} - x^+ - x^-)\log\left(\frac{ t_{\infty} - x^{-} }{ t_{\infty} } \right)
        \right].
\end{aligned}
\end{equation}
Now, plugging this back into the $+-$ component of the dilaton equations of motion, which reads
\begin{equation}
\begin{aligned}
    \partial_{x^+ x^-}\phi + \frac{ 2\phi }{ (x^+ - x^-)^2  } = 0,
\end{aligned}
\end{equation}
we find that we must have
\begin{equation}
\begin{aligned}
    B = \frac{\mathcal{C} - k }{ 2 }.
\end{aligned}
\end{equation}
giving finally \eqref{eq:phiDataLate}. As a double check, we find that when plugging \eqref{eq:phiDataLate} back into
the dilaton equations of motion, we get that the solution is indeed sourced by \eqref{eq:Tmmcont}.

\subsection*{Transforming to global coordinates}
\eqref{eq:globalcoords} and some basic algebra gives\footnote{Solving for $x^{\pm}(w^{\pm})$ from \eqref{eq:globalcoords} also gives another
branch, but with our chosen value of $\tau_0$ the given one will be the one where $x^{\pm}(w^{\pm})$ is continuous in the right Poincare
patch.}
\begin{align}
    x^{\pm} &= -\lambda + \mu^{-1} f_{\pm}(w^{\pm}), \qquad
    f_{\pm}(z) \equiv \tan(z-\tau_0) \pm \frac{ 1 }{ \cos(z-\tau_0) }, 
    \label{eq:xtow}\\
    \frac{ \partial w^{\pm} }{ \partial x^{\pm} } &= 
    \mu\left[1 \mp \sin(w^{\pm}-\tau_0)\right] \label{eq:jacobian}.
\end{align}

As indicated in the main text, we want to solve 
\begin{equation}\label{eq:isocond}
\begin{aligned}
    \tau(z_*, t_*) &= \tau(0, t_{\partial}) = \rho(z_*, t_*) = 0, \\
\end{aligned}
\end{equation}
for $\lambda, \mu, \tau_0$, where $(t_*, z_*)$ is the QES position in Poincare coordinates. 

Through \eqref{eq:globalcoords}, \eqref{eq:isocond} translates to
\begin{equation} \label{eq:consteqs}
\begin{aligned}
    (t_* + \lambda)\mu &= -\tan \tau_0, \\
    [f(u_{\partial})+\lambda] \mu &= - \frac{ \sin \tau_0 }{ 1 + \cos \tau_0  }, \\
    z_* \mu &= \frac{ 1 }{ \cos \tau_0 }.
\end{aligned}
\end{equation}
Next, we know the QES location to leading order is given by
\begin{equation}\label{eq:QESloc}
\begin{aligned}
    x^+_* &= t_{\infty} + \frac{ 1 }{ 3 }\left[t_{\infty} -
    f(u_{\partial})\right], \\
    x^-_* &= t_{\infty} - \frac{ 8\pi T_1 e^{-kf^{-1}(x^-_*)/2 } }{ 3k
    }\left[ t_{\infty} - f(u_{\partial})\right].
\end{aligned}
\end{equation}
Solving now \eqref{eq:consteqs} and inserting \eqref{eq:QESloc}, we find 
\begin{align}
    \chi & \equiv \mu(t_{\infty}+\lambda) = -\frac{ 1 }{ 4 } 
    + \frac{ 5 e^{-k f^{-1}(x^-_*)/2 } k }{ 8\pi T_1  } 
    + \mathcal{O}(k^2 ),  \\
    \tau_0 &= \frac{ \pi }{ 2 } - \frac{ k e^{-kf^{-1}(x^-_*)}  }{ \pi T_1 } + \mathcal{O}(k^2),
     \\
    \mu &= \frac{ 1 }{ t_{\infty} - f(u_{\partial})}\left(\frac{ 3 }{ 4 } -
    \frac{ 3k e^{kf^{-1}(x^-_*)/2} }{  8 \pi T_1 } + \mathcal{O}(k^2)
    \right).
\end{align}
Next, from \cite{AEMM} we have at the QES that 
\begin{equation}
\begin{aligned}
    f^{-1}(x^{-}_*(u)) \approx u - \frac{ e^{ku/2} }{ 2\pi T_1  }\log\left(\frac{ 8\pi T_1 e^{-ku/2}}{ 3k } \right),
\end{aligned}
\end{equation}
which gives that
\begin{equation}
\begin{aligned}
    e^{kf^{-1}(x^{-1}_*)/2} =  e^{k u_{\partial} / 2}\left[1 + \mathcal{O}(k \log k
    )\right],
\end{aligned}
\end{equation}
giving finally \eqref{eq:chidef}, \eqref{eq:tau0sol} and \eqref{eq:musol}.

\subsection*{The stress-tensor initial boundary value problem}
Consider general initial data for the CFT stress tensor in global AdS on the $\tau=0$ slice:
\begin{equation}
\begin{aligned}
    T_{w^- w^-}|_{\tau=0} &= g(\rho) = g(w^-)|_{\Sigma}, \\
    T_{w^+ w^+}|_{\tau=0} &= h(-\rho) = h(w^+)|_{\Sigma}.
\end{aligned}
\end{equation}
Before we even specify boundary conditions, the fact that solutions $T_{w^-w^-}, T_{w^+w^+}$ are
functions of only $w^-$ and $w^+$, respectively, imposes the following boundary values
\begin{equation}\label{eq:bcinduced}
\begin{aligned}
    T_{w^- w^-}|_{\partial M_R} &= g(w^-)=g\left(\tau + \frac{ \pi }{ 2 } \right), \qquad \tau\in \left[- \frac{ \pi }{
        2 }, 0\right], \\
    T_{w^- w^-}|_{\partial M_L} &= g(w^-)=g\left(\tau - \frac{ \pi }{ 2 } \right), \qquad \tau\in \left[0, \frac{ \pi }{ 2
    }\right], \\
    T_{w^+ w^+}|_{\partial M_R} &= h(w^+)=h\left(\tau - \frac{ \pi }{ 2 } \right), \qquad \tau\in \left[0, \frac{ \pi }{ 2
    }\right], \\
    T_{w^+ w^+}|_{\partial M_L} &= h(w^+)=h\left(\tau + \frac{ \pi }{ 2 } \right), \qquad \tau\in \left[-\frac{ \pi }{
        2 }, 0\right],
\end{aligned}
\end{equation}
where $\partial M_{L/R}$ are the left/right conformal boundaries.
This is obtained simply by tracing null rays from the initial data to the boundary, holding the relevant stress-tensor
component fixed. Now let us consider reflection boundary conditions
\begin{equation}\label{eq:bcfixed}
\begin{aligned}
    \left(T_{w^-w^-} - T_{w^+w^+}\right)|_{\partial M} &= 0.
\end{aligned}
\end{equation}
The combination of \eqref{eq:bcinduced} and \eqref{eq:bcfixed} now determines the boundary values of $T_{w^{\pm}w^{\pm}}$
for all $\tau \in \left[-\frac{ \pi }{ 2 }, \frac{ \pi }{ 2 }\right]$:
\begin{equation}\label{eq:completeBCs}
\begin{aligned}
    T_{w^-w^-}|_{\partial M_R} &= T_{w^+w^+}|_{\partial M_R} = \theta(-\tau) g\left(\tau + \frac{ \pi }{ 2
    }\right)+\theta(\tau)h\left(\tau - \frac{ \pi }{ 2 }\right) + b_R\delta(\tau),  \\
    T_{w^-w^-}|_{\partial M_L} &= T_{w^+w^+}|_{\partial M_L} = \theta(\tau)g\left(\tau - \frac{ \pi }{ 2 } \right) +
    \theta(-\tau)
    h\left(\tau + \frac{ \pi }{ 2 }\right) + b_L \delta(\tau),
\end{aligned}
\end{equation}
Note that we have discontinuities at $\tau=0$, which stems from the fact that our initial data does not satisfy
\eqref{eq:bcfixed} at $\tau=0$ for general $g$ and $h$. Due to the discontinuous stress tensors at $\tau=0$ there is 
no way to rule out a $\delta$-function shock at $\tau=0$, and so this must be allowed. Altering the boundary conditions
in AdS causes the injection of energy \cite{AEMM}, and since we are considering the scenario where we turn off absorbing
boundary condition very fast to the future and the past, we will have that $b_R, b_L$ are nonzero. 
The specific value of $b_L$ and $b_R$ will depend on the specifics of the theory and the state, and 
so they cannot be determined at our level of analysis. However, in order to have that the spacetime energy is constant, we
must have that that the strength $w^+$ and $w^-$ shocks on each boundary is the same, which we used in
\eqref{eq:completeBCs}.

Rewriting these expressions in terms of $w^{\pm}$ we finally obtain the full solution for all $|w^{\pm}|\leq \pi$:
\begin{equation}\label{eq:completeTsol}
\begin{aligned}
    T_{w^-w^-} &= \theta\left(-w^{-}+ \frac{ \pi }{ 2 } \right) g\left(w^-\right)+\theta\left( 
    w^-- \frac{ \pi }{ 2 }\right)h\left(w^- - \pi\right) + b_R \delta\left(w^- - \frac{ \pi }{ 2 } \right), \qquad w^{-} \in \left[0, \pi \right] \\
    T_{w^-w^-} &= \theta\left(w^- + \frac{ \pi }{ 2 } \right)g\left(w^{-}\right) + \theta\left(-w^{-} - \frac{ \pi }{ 2
    } \right) h\left(w^- + \pi \right) + b_L \delta\left(w^- + \frac{ \pi }{ 2 } \right), \qquad w^{-} \in \left[-\pi, 0 \right] \\
    T_{w^+w^+} &= \theta\left(w^+ + \frac{ \pi }{ 2 }\right)h\left(w^+\right) + \theta\left( -w^+ - \frac{ \pi }{ 2 }\right)\
     g\left(w^+ + \pi \right) + b_R \delta\left(w^+ + \frac{ \pi }{ 2 } \right), \qquad w^{+} \in \left[-\pi, 0 \right] \\
    T_{w^+w^+} &= \theta\left(-w^+ + \frac{ \pi }{ 2 }\right)h\left(w^+\right) + \theta\left(w^+ - \frac{ \pi }{ 2 }\right)
     g\left(w^+ - \pi\right) + b_L \delta\left(w^+ - \frac{ \pi }{ 2 } \right),  \qquad w^{+} \in \left[0,  \pi\right].
\end{aligned}
\end{equation}
In the domain covered by $|w^{\pm}|\leq \pi$, this represents the distributional evolution that agrees (1) agrees with
our initial data everywhere in the domain of the dependence, and (2) has reflecting boundary conditions in the weak
sense.
Note also that in the case relevant for us, where $\Sigma$ has CPT symmetry about the QES, we must have $b_L = b_R$.
Combining \eqref{eq:initialdataT} with \eqref{eq:completeTsol} gives \eqref{eq:Tevolution}.

\subsection*{Evolving the dilaton}
Thanks to the CPT symmetry, it is sufficient to restrict to the region in the future and spacelike to the right of the
QES, which is covered by $w^- \geq 0$. Thus, in what follows we always work in the range $0 \leq w^-\leq \pi$.

Changing to global coordinates in \eqref{eq:phiDataLate} gives the
dilaton in $D[\Sigma]$ (i.e.\ for $w^+ < 0, w^- > 0$) on the form:
\begin{equation}\label{eq:phiGlobalInit}
\begin{aligned}
    \phi(w^+, w^-)|_{D[\Sigma]} &= 
    \frac{ \bar{\phi}_r }{ 2(f_+ - f_-) }\Big[ (2\chi - f_{-}-f_+)\mathcal{C}  -k(f_+ - f_-) \\
    &\qquad \qquad\qquad\qquad  + k(2\chi - f_+ - f_-) \log\left(\frac{\chi - f_{-}}{t_{\infty}\mu}\right)
    \Big],  
\end{aligned}
\end{equation}
Now we compute $\hat{I}^{-}(w^+, w^-)$ for $0<w^-<\frac{ \pi }{ 2 }$.
The integral can be done analytically and reads
\begin{equation}
\begin{aligned}
    8\pi G_N \hat{I}^{-}(w^+, w^-) &= -8\pi G_N \int_{\varepsilon}^{w^-} \dd s \left[1+\sin(s-\tau_0)\right]\left[f_-(s) - f_+(w^+) \right]\left[f_-(s) - f_-(w^-)
    \right]g(s) \\
    &= \frac{ 1 }{ 2 }k \bar{\phi}_r\left[\frac{ 2 }{ k \delta \tau_0 } + (\chi+ f_- - f_+) + (2\chi - f_+ -
    f_-)\log\left(\frac{ (\chi - f_-)k\delta \tau_0 }{ 2 } \right) \right]  \\
    &\equiv H_1(w^+, w^-), \label{eq:H1}
\end{aligned}
\end{equation}
where we introduced the notation
\begin{equation}
\begin{aligned}
    \tau_0 = \frac{ \pi }{ 2 } - k \delta \tau_0 + \mathcal{O}(k^2). 
\end{aligned}
\end{equation}
$\hat{I}^+$ vanishes on $D[\Sigma]$, and so we can write \eqref{eq:phiGlobalInit} 
\begin{equation}
\begin{aligned}
    \phi &= \phi_{0} + \phi_{m}, \\
    \phi_{m} &\equiv \frac{  H_1(w^+, w^-) }{ f_+ - f_- }, \\
    \phi_0 &\equiv \phi - \phi_{m} =  \frac{ c_1^{(0)} + c_2^{(0)}(f_+ + f_-) + c_3^{(0)} f_+ f_-  }{ f_+ - f_- },
\end{aligned}
\end{equation}
with
\begin{equation}
    \begin{aligned}\label{eq:cinit}
        c_1^{(0)} &= \bar{\phi}_{r}\left[\mathcal{C}\chi - \frac{ 1 }{ \delta \tau_0 } - \frac{ k }{ 2 }\chi - k\chi \log\left(\frac{
        k \delta \tau_0 \mu }{ 2 } \right)\right], \\
    c_2^{(0)} &= -\frac{ 1 }{ 2 }\bar{\phi}_r \left[\mathcal{C}-k\log\left(\frac{
        k \delta \tau_0 \mu }{ 2 } \right)  \right], \\
    c_3^{(0)} &= 0. \\
\end{aligned}
\end{equation}

Now let us evolve out of $D[\Sigma]$. Let's focus first on the part of the dilaton sourced
by continuous stress-energy, focusing on the $\delta$-function contributions afterwards. We can ignore the $\varepsilon$ for
this, as they are only there to make sure we are not lying exactly on top of a shock.

Thanks to the step-function in \eqref{eq:Tevolution}, $\hat{I}^-$ for $\frac{ \pi }{ 2 }< w^-< \pi$:
\begin{equation}
\begin{aligned}
    8\pi G_N \hat{I}^-|_{\rm cont} &=  -8\pi G_N \int_{0}^{\pi/2} \dd s \left[1+\sin(s-\tau_0)\right]\left[f_-(s) - f_+(w^+) \right]\left[f_-(s) - f_-(w^-)
    \right]g(s)  \\
    &= \frac{ k\bar{\phi}_r }{ 2 }\Big[ 
    \frac{ 2 }{ k\delta \tau_0 } + \frac{-1 + f_+ f_- -(1+f_+ + f_-)\chi + \chi^2 }{ 1+\chi } \\
    &\qquad \qquad \qquad \qquad+ (2\chi - f_+ - f_-)
    \log\left(\frac{ k \delta \tau_0 (1+\chi) }{ 2 } \right) 
    \Big] \\
    &= H_2(w^+, w^-). \label{eq:H2}
\end{aligned}
\end{equation}

Next lets turn to the continuous part of $\hat{I}^+$, which becomes nonzero either when $\hat{w}^+ \geq \frac{ \pi }{ 2
}$ or $\hat{w}^+ \leq -\frac{ \pi }{ 2 }$. Consider first the latter range. We find, ignoring $\mathcal{O}(k^2)$
corrections,
\begin{equation}
\begin{aligned}
    8\pi G_N \hat{I}^{+}|_{\rm cont} &=
    8\pi G_N \int_{-\frac{ \pi }{ 2 }}^{w^+} \dd s \left[1-\sin(s-\tau_0)\right]\left[f_+(s) - f_+(w^+) \right]\left[f_+(s) -
    f_-(w^-)
    \right]g(s + \pi) \\
    &= \frac{ \bar{\phi}_r k }{ 2 }\left[ (f_+ + f_- - 2\chi)\log\left|\chi - \tan \frac{ s }{ 2 } \right| - \frac{ (f_-
    - \chi)(f_+ - \chi) }{ \chi\left(\chi \cos \frac{ s }{ 2 } - \sin \frac{ s }{ 2 } \right)  } - \tan \frac{ s }{ 2
    }\right] \Bigg|_{s=-\frac{ \pi }{ 2 }}^{s= w^+} \\ 
    & \equiv H_{3}(w^+, w^-). \label{eq:H3}
\end{aligned}
\end{equation}
Next, in the range $w^+ \geq \frac{ \pi }{ 2 }$ we find 
\begin{equation}
\begin{aligned}
    \hat{I}^{+}|_{\rm cont} &=
    \int_{\frac{ \pi }{ 2 }}^{w^+} \dd s \left[1-\sin(s-\tau_0)\right]\left[f_+(s) - f_+(w^+) \right]\left[f_+(s) -
    f_-(w^-)
    \right]g(s - \pi) \\
    &= \frac{ \bar{\phi}_r k }{ 2 }\left[ (f_+ + f_- - 2\chi)\log\left|\chi-\tan \frac{ s }{ 2 } \right| - \frac{ (f_-
    - \chi)(f_+ - \chi) }{ \chi\left(\chi \cos \frac{ s }{ 2 } - \sin \frac{ s }{ 2 } \right)  } - \tan \frac{ s }{ 2
    }\right] \Bigg|_{s=\frac{ \pi }{ 2 }}^{s= w^+} \\ 
    & \equiv H_{4}(w^+, w^-). \label{eq:H4}
\end{aligned}
\end{equation}

Next we turn to the contribution of the shocks. Let us start with $\hat{I}^-$. We get a jump as we cross $w^- = \frac{
    \pi }{ 2 }$:
\begin{equation}
\begin{aligned}
    8\pi G_N \hat{I}^-|_{\rm shock} &= 
    -8 \pi G_N \int_{\epsilon}^{w^-} \dd s \left[1+\sin(s-\tau_0)\right]\left[f_-(s) - f_+(w^+) \right]\left[f_-(s) - f_-(w^-)
    \right]b\delta\left(s-\frac{ \pi }{ 2 }\right) \\
    &= - 8\pi G_N b\theta\left( w^- -\frac{ \pi }{ 2 }\right)(1+f_+)(1+f_{-}), \label{eq:Iminshock}
\end{aligned}
\end{equation}
where we neglect non-perturbatively small corrections. Next, consider $w^+$. Here we have three ranges of interest.
By a similar computation as above we have
\begin{align}
    \hat{I}^+\left(w^+ < - \frac{ \pi }{ 2 }\right)|_{\rm shock} 
    &= 8\pi G_N b (1+f_+)(1+f_-), \label{eq:IplushockA}\\
    \hat{I}^+\left( 0 < w^+ < \frac{ \pi }{ 2 }\right)|_{\rm shock} 
    &= 16\pi G_N a_+ f_+ f_- , \label{eq:IplusshockB}\\
    \hat{I}^+\left( \frac{ \pi }{ 2 } < w^+ < \pi \right)|_{\rm shock}
    &= 16\pi G_N a_+ f_+ f_-+8\pi G_N b(f_+ -1)(f_- - 1) \label{eq:IplushockC}
\end{align}
These contributions to $\phi$ can most conveniently be included by modifying the $c_1, c_2, c_3$ coefficients from
earlier to be step functions:
\begin{equation}
\begin{aligned}
    c_1 &= c_1^{(0)} + 8\pi G_N b \left[-\theta\left(w^{-} - \frac{ \pi }{ 2 } \right) + \theta\left(-\frac{ \pi }{ 2
    } + w^+ \right) + \theta\left(w^+ - \frac{ \pi }{ 2 } \right) \right], \\
    c_2 &= c_2^{(0)} + 8\pi G_N b \left[-\theta\left(w^{-} - \frac{ \pi }{ 2 } \right)+ \theta\left(-\frac{ \pi }{ 2
    } + w^+ \right) -\theta\left(w^+ - \frac{ \pi }{ 2 } \right) \right], \\
    c_3 &= c_3^{(0)}+ 8\pi G_N b \left[-\theta\left(w^{-} - \frac{ \pi }{ 2 } \right)+ \theta\left(-\frac{ \pi }{ 2
    } + w^+ \right) +\theta\left(w^+ - \frac{ \pi }{ 2 } \right)\right] + 16 \pi G_N a_+
    \theta(w^+).
\end{aligned}
\end{equation}
This completes the determination of the dilaton. 

\subsection*{Bounding $b$}
Let us now find the leading order bound on $b$ by demanding that the QES is causally disconnected from the conformal
boundary. This means that we need that the boundary endpoints of the physical conformal boundary lies in the interval $|\tau| \leq \frac{ \pi }{ 2 }$. Consider the future
boundary, which we thus require to lie in the region $\frac{ \pi }{ 2 } \leq w^-$, $-\pi <w^+ < \frac{ \pi }{ 2 }$. 
Evaluating the dilaton on the boundary where $f_+ = f_-$, we get to leading order that
\begin{equation}
\begin{aligned}
    (f_+ - f_-)\phi|_{\partial M_R} &= c_1 + 2 c_2 f_+ + c_3 f_+^2 + H_2 \\
    &= -8\pi G_N (1+f_+)^2 - 4\pi T_1 \bar{\phi}_r e^{-u_{\partial}/2}f_+ + \mathcal{O}(k\log k),
\end{aligned}
\end{equation}
where $w^+ = \tau - \frac{ \pi }{ 2 }$. In order for the physical boundary to terminate in the future some $\tau \leq \pi/2$ we need
\begin{equation}
\begin{aligned}
    -8\pi G_N b \left[ 1+f_+\left(\tau - \pi/2 \right)\right]^2 - 4 t_{\infty}^{-1}\bar{\phi}_r
    e^{-u_{\partial}/2}f_+\left(\tau - \pi/2 \right) = 0
\end{aligned}
\end{equation}
has a solution in the range $0<\tau<\frac{ \pi }{ 2 }$. The above can be simplified to 
\begin{equation}
\begin{aligned}
    \tan^2 \tau = \frac{ c e^{-u_{\partial}k} T_1 }{ 6 \pi b^2 k^2 t_{\infty} }\left(b e^{u_{\partial}k/2}k + \frac{ c
    }{ 24\pi t_{\infty}} \right).
\end{aligned}
\end{equation}
This only has a solution if 
\begin{equation}
\begin{aligned}
b > - \frac{ c }{ 24\pi }\frac{ e^{- u_{\partial}k/2}}{t_{\infty} k} = -2a_-.
\end{aligned}
\end{equation}
A similar computation for $\tau<0$ gives the slightly stronger bound
\begin{equation}
b > - \frac{ c }{ 30\pi }\frac{ e^{- u_{\partial}k/2}}{t_{\infty} k}.
\end{equation}

\subsection{Computing the Pre Page Time Canonical Purification in JT Gravity}
Let us now build the canonical purification before the Page time, purifying at some specific boundary time $u=u_{\partial}$. For analytical convenience we work at times of order
$u_{\partial} \sim \mathcal{O}(1)$, but the general picture is also valid shortly before the Page time.

The stress tensor on the domain of dependence for our initial data slice $\Sigma$ is given by \eqref{eq:TAEMM}. 
Since we do not keep track of $O(k^2)$ corrections, we can use the early time expression
\begin{equation}
\begin{aligned}
    f(u) = \frac{ 1 }{ \pi T_1 }\tanh\left(\pi T_1 u \right). \label{eq:fearly}
\end{aligned}
\end{equation}
We can see this by noting that (1) the expression $\hat{f}(u) \equiv (\pi T_1)^{-1}\tanh\left[\frac{ 2\pi T_1 }{ k }(1-
e^{-ku/2})\right]$ solves the differential equation for $f(u)$ to order $O(k^2)$,\footnote{The differential equation
determining $f(u)$ is \eqref{eq:Eschw}=\eqref{eq:Esol}.}
and (2) up to $O(k^2)$ corrections, \eqref{eq:fearly} and $\hat{f}(u)$ gives the same value for $\{f^{-1}(x), x\}$. 
Inverting for $f^{-1}$ and computing the Schwarzian derivative, we get \eqref{eq:Tmmearly}
\begin{equation}
\begin{aligned}
    T_{x^-x^-}|_{D[\Sigma]} = E_S \delta(x^-) - \frac{ c }{ 24\pi }\theta(x^-) \frac{ 2(\pi T_1)^2 }{ \left[1 - (\pi T_1
    x^-)^2\right]^2 } \equiv h(x^-).
\end{aligned}
\end{equation}

We want to canonically purify the state at $t=t_{\partial}=f(u_{\partial})$. The fact that $T_{x^{\pm}x^{\pm}}$ is constant along lines of
constant $x^{\pm}$ implies that 
\begin{equation}
\begin{aligned}
    T_{x^-x^-}(t_1 < t < t_{\partial})|_{\partial M}  &= h(t), \\ 
    T_{x^+x^+}(t_{\partial}< t < t_2)|_{\partial M}  &= 0, \\ 
\end{aligned}
\end{equation}
where $t_2(/t_1)$ is the future-most (/past-most) boundary time in causal contact with $\Sigma$.
Imposing reflecting boundary conditions gives
\begin{equation}
\begin{aligned}
    T_{x^-x^-}|_{\partial M} = T_{x^+x^+}|_{\partial M} = \theta(t_{\partial}-t)h(t) + b \delta(t-t_{\partial}),
    \qquad t\in[t_1, t_2],
\end{aligned}
\end{equation}
which fixes the solution
\begin{equation}
\begin{aligned}
    T_{x^-x^+} &= \theta(t_{\partial}-x^-)h(x^-) + b\delta(x^- - t_{\partial}),\\
    T_{x^+x^+} &= \theta(t_{\partial}-x^+)h(x^+) + b\delta(x^+ - t_{\partial}),
\end{aligned}
\end{equation}
where Fig.~X indicates the domain where this solution is fixed.
As was the case after the Page time, $b$ is the amplitude of the shock caused by turning of absorbing boundary conditions.

Now we compute the dilaton. Choosing our reference point for the stress tensor integrals at $x^{\pm} = t_{\partial}\pm
\varepsilon$ for some small positive $\varepsilon$, we have
\begin{equation}
\begin{aligned}
    I^+(x^+, x^-) 
    &= \int_{t_{\partial} + \varepsilon}^{x^+}\dd s(s-x^+)(s-x^-)\left[ \theta(t_{\partial}-s)h(s) + b\delta(s - t_{\partial}) \right] \\
    &=  b(t_{\partial}-x^+)(t_{\partial}-x^-)\theta(x^+ - t_{\partial})
    -\theta(-x^+)x^+ x^- E_S
    - \theta(t_{\partial}- x^+)F(x^+)
\end{aligned}
\end{equation}
where 
\begin{equation}\label{eq:Fdef}
\begin{aligned}
    F(x^+) &= -\frac{ c }{ 24\pi }\int_{\max\{x^+, 0\}}^{t_{\partial}}\dd s (s-x^+)(s-x^-)\theta(s) \frac{ 2(\pi T_1)^2 }{ \left[1 - (\pi T_1
    s)^2\right]^2 } \\
    &= - \frac{ c }{ 24\pi }\left[\frac{ x^+ + x^- - s\left(1+ \pi^2 T_1^2 x^+ x^- \right)}{ \pi^2 T_1^2 s^2 -1 } +
    \left(\pi T_1 x^+ x^+ -\frac{ 1 }{ \pi T_1 } \right)\arctan\left(\pi T_1 s\right)\right]\Big|_{s= \max\{x^+,
    0\}}^{s=t_{\partial}}.
\end{aligned}
\end{equation}
Similarly we get
\begin{equation}
\begin{aligned}
    I^-(x^+, x^-) &= - b\theta(x^- - t_{\partial})(t_{\partial}-x^+)(t_{\partial}-x^-) + \theta(-x^-) E_S x^+ x^-
    + \theta(t_{\partial}-x^-)F(x^-).
\end{aligned}
\end{equation} 
In total, this gives a dilaton
\begin{equation}
\begin{aligned}
    \phi &= \frac{ c_1 + c_2(x^+ + x^-) + c_3 x^+ x^- }{ x^+ - x^- } + \phi_m , \\
    \phi_m &=  8\pi G_N\frac{\theta(t_{\partial}-x^-)F(x^-)-\theta(t_{\partial}-x^+)F(x^+) }{ x^+ - x^- },
\end{aligned}
\end{equation}
with the piecewise constant coefficients $c_{i}$ 
\begin{equation}
\begin{aligned}\label{eq:cdef}
    c_1 &=  2 \bar{\phi}_r + 8\pi G_N b t_{\partial}^2\left[\theta(x^+ - t_{\partial})-\theta(x^- - t_{\partial})
    \right] \\
    c_2 &= -8\pi G_N t_{\partial}\left[\theta(x^+ - t_{\partial})-\theta(x^- - t_{\partial})
    \right] \\
    c_3 &= -2\bar{\phi}_r(\pi T_1)^2 + 8\pi G_N \left[b \theta(x^+ - t_{\partial})-b \theta(x^- - t_{\partial}) - E_S
    \theta(-x^+) + E_S \theta(-x^-) ) 
    \right], 
\end{aligned}
\end{equation}
where the coefficients in $D[\Sigma]$ is fixed by the knowledge that the homogeneous part of the solution at our reference point is that of \eqref{eq:phiBH} with temperature $T=T_1$.

\bibliographystyle{jhep}
\bibliography{all}
\end{document}